\documentclass[3p,,preprint,12pt]{elsarticle}
\makeatletter\if@twocolumn\PassOptionsToPackage{switch}{lineno}\else\fi\makeatother

\usepackage{algorithm}%
\usepackage{algorithmicx}%
\usepackage{algpseudocode}
\usepackage{hyperref}

\usepackage{tabulary,xcolor}
\usepackage{amsfonts,amsmath,amssymb}
\usepackage[T1]{fontenc}
\usepackage{booktabs}
\usepackage[inline]{enumitem}
\usepackage{stmaryrd,wrapfig}
\usepackage{textcmds,cleveref}
\usepackage{subcaption}
\usepackage{tabularx}
\usepackage{multicol}
\usepackage{array}

\makeatletter
\let\save@ps@pprintTitle\ps@pprintTitle
\def\ps@pprintTitle{\save@ps@pprintTitle\gdef\@oddfoot{\footnotesize\itshape \null\hfill}}
\def\hlinewd#1{%
  \noalign{\ifnum0=`}\fi\hrule \@height #1%
  \futurelet\reserved@a\@xhline}

\AtBeginDocument{\ifNAT@numbers \biboptions{sort&compress}\fi}

\makeatother

  \renewenvironment{abstract}{\global\setbox\absbox=\vbox\bgroup
    \hsize=\textwidth%
  \noindent\unskip\textbf{}
   \par\medskip\noindent\unskip\ignorespaces}
   {\egroup}
  
\usepackage{ifluatex}
\ifluatex
\usepackage{fontspec}
\defaultfontfeatures{Ligatures=TeX}
\usepackage[]{unicode-math}
\unimathsetup{math-style=TeX}
\else 
\usepackage[utf8]{inputenc}
\fi 
\ifluatex\else\usepackage{stmaryrd}\fi

\usepackage{url,multirow,morefloats,floatflt,cancel,tfrupee}
\makeatletter

\AtBeginDocument{\@ifpackageloaded{textcomp}{}{\usepackage{textcomp}}}
\makeatother
\usepackage{colortbl}
\usepackage{xcolor}
\usepackage{pifont}
\usepackage[nointegrals]{wasysym}
\makeatletter

\def\mcWidth#1{\csname TY@F#1\endcsname+\tabcolsep}

\def\cAlignHack{\rightskip\@flushglue\leftskip\@flushglue\parindent\z@\parfillskip\z@skip}
\def\rAlignHack{\rightskip\z@skip\leftskip\@flushglue \parindent\z@\parfillskip\z@skip}

\@ifundefined{etal}{}{}

\usepackage{ifxetex}
\ifxetex\else\if@twocolumn\@ifpackageloaded{stfloats}{}{\usepackage{dblfloatfix}}\fi\fi

\AtBeginDocument{
\expandafter\ifx\csname eqalign\endcsname\relax
\def\eqalign#1{\null\vcenter{\def\\{\cr}\openup\jot\m@th
  \ialign{\strut$\displaystyle{##}$\hfil&$\displaystyle{{}##}$\hfil
      \crcr#1\crcr}}\,}
\fi
}

\AtBeginDocument{%
  \@ifpackageloaded{endfloat}%
   {\renewcommand\efloat@iwrite[1]{\immediate\expandafter\protected@write\csname efloat@post#1\endcsname{}}}{\newif\ifefloat@tables}%
}%

\def\BreakURLText#1{\@tfor\brk@tempa:=#1\do{\brk@tempa\hskip0pt}}
\let\lt=<
\let\gt=>
\def\processVert{\ifmmode|\else\textbar\fi}

\@ifundefined{subparagraph}{
\def\subparagraph{\@startsection{paragraph}{5}{2\parindent}{0ex plus 0.1ex minus 0.1ex}%
{0ex}{\normalfont\small\itshape}}%
}{}

\newcommand\role[1]{\unskip}
\newcommand\aucollab[1]{\unskip}
  
\@ifundefined{tsGraphicsScaleX}{\gdef\tsGraphicsScaleX{1}}{}
\@ifundefined{tsGraphicsScaleY}{\gdef\tsGraphicsScaleY{.9}}{}
\def\checkGraphicsWidth{\ifdim\Gin@nat@width>\linewidth
	\tsGraphicsScaleX\linewidth\else\Gin@nat@width\fi}

\def\checkGraphicsHeight{\ifdim\Gin@nat@height>.9\textheight
	\tsGraphicsScaleY\textheight\else\Gin@nat@height\fi}

\def\fixFloatSize#1{}
\let\ts@includegraphics\includegraphics

\def\inlinegraphic[#1]#2{{\edef\@tempa{#1}\edef\baseline@shift{\ifx\@tempa\@empty0\else#1\fi}\edef\tempZ{\the\numexpr(\numexpr(\baseline@shift*\f@size/100))}\protect\raisebox{\tempZ pt}{\ts@includegraphics{#2}}}}

\AtBeginDocument{\def\includegraphics{\@ifnextchar[{\ts@includegraphics}{\ts@includegraphics[width=\checkGraphicsWidth,height=\checkGraphicsHeight,keepaspectratio]}}}

\DeclareMathAlphabet{\mathpzc}{OT1}{pzc}{m}{it}

\def\URL#1#2{\@ifundefined{href}{#2}{\href{#1}{#2}}}

\def\UrlOrds{\do\*\do\-\do\~\do\'\do\"\do\-}%
\g@addto@macro{\UrlBreaks}{\UrlOrds}

\edef\fntEncoding{\f@encoding}

\makeatother

\newif\ifmultipleabstract\multipleabstractfalse%
    \makeatletter
\def\ead{\@ifnextchar[{\@uad}{\@ead}}
\gdef\@ead#1{\bgroup
   \def\_{\string\underscorechar\space}
   \def\{{\string\lbracechar\space}
   \def\textdagger{\string\textdagger\space}
   \def\texttildeapprox{\string\texttildeapprox\space}
   \def~{\hashchar\space}
   \def\}{\string\rbracechar\space}
   \edef\tmp{\the\@eadauthor}
   \immediate\write\@auxout{\string\emailauthor
     {#1}{\expandafter\strip@prefix\meaning\tmp}}
  \egroup
}
\gdef\emailauthor#1#2{\stepcounter{ead}
      \g@addto@macro\@elseads{\raggedright
      \let\corref\@gobble
      \eadsep\texttt{#1} (#2)
      \def\eadsep{\unskip,\space}}
}

\makeatother
  
\begin{document}

\begin{frontmatter}

\title{Dynamic Modeling and Load-Following Control of Small Modular Reactors with Moving-Boundary Steam Generators and Thermodynamically Coupled Rankine Cycles}
 
\author[1,2]{Ali Mahboub Rad}
\author[2]{Roshni Anna Jacob}
\author[3]{Bikash Poudel}
\author[4]{Mayir Mamtimin}
\author[1,2,5]{Jie Zhang}

\address[1]{Department of Electrical and Computer Engineering, The University of Texas at Dallas, Richardson, Texas 75080, USA}
\address[2]{Department of Mechanical Engineering, The University of Texas at Dallas, Richardson, Texas 75080, USA}
\address[3]{Idaho National Laboratory, Idaho Falls, Idaho 83415, USA}
\address[4]{American Bureau of Shipping, Spring, Texas 77389, USA}
\address[5]{Corresponding author}
    
\begin{abstract}
\textbf{Abstract:} Small modular reactors (SMRs) are increasingly considered for flexible power generation; however, many dynamic studies still neglect the thermodynamic coupling between the primary and secondary loops that is essential for accurate assessment of load-following capability. In this study, we develop a hybrid dynamic framework that couples an equation-based model of a NuScale-type integral pressurized water reactor, including the reactor, primary loop, and moving-boundary helical-coil once-through steam generator, with a physics-based secondary Rankine cycle comprising the steam throttle valve, turbine, condenser, and feedwater pump. This approach enforces mass and energy conservation across the coupled system while preserving physically consistent pressure--flow and enthalpy--flow interactions across the domain boundary. The integrated model reproduces nominal design-point conditions and is used to analyze a 5\% step reduction in turbine mechanical-power demand under five control configurations, including a decentralized three-loop control architecture for the valve, feedwater pump, and control rods. The results show that partial control strategies can satisfy individual objectives but leave pressure, thermal, or phase-boundary deviations, whereas simultaneous action of all three actuators provides the most balanced response by stabilizing steam pressure, limiting primary-loop thermal deviations, and maintaining acceptable steam-generator operating margins during load-following maneuvers. Compared with a conventional linear steam-cycle representation, the coupled framework captures dynamic back-pressure and variable turbine enthalpy drop that are otherwise neglected, leading to different predictions of transient behavior and required steam flow. These findings show that thermodynamically coupled, physics-based steam-cycle models are needed for more accurate assessment of SMR operational flexibility and operating margins under realistic load-following conditions.
\end{abstract}

    
\end{frontmatter}

\begingroup
\setlength{\fboxsep}{8pt}
\setlength{\fboxrule}{0.4pt}

\noindent\fbox{%
\begin{minipage}{\dimexpr\textwidth-2\fboxsep-2\fboxrule\relax}
\scriptsize

\textbf{Nomenclature}

\vspace{0.5em}

\setlength{\columnsep}{1.0cm}
\begin{multicols}{2}

\textit{Abbreviations}
\begin{description}[
    style=sameline,
    leftmargin=1.65cm,
    labelwidth=1.35cm,
    labelsep=0.3cm,
    itemsep=0pt,
    parsep=0pt,
    topsep=2pt
]
\item[BOP] balance of plant
\item[HP] high pressure
\item[iPWR] integral pressurized water reactor
\item[LP] low pressure
\item[PI] proportional--integral
\item[SG] steam generator
\item[SMR] small modular reactor
\end{description}

\vspace{0.4em}

\textit{Symbols}
\begin{description}[
    style=sameline,
    leftmargin=1.65cm,
    labelwidth=1.35cm,
    labelsep=0.3cm,
    itemsep=0pt,
    parsep=0pt,
    topsep=2pt
]
\item[\(A\)] area, \(\mathrm{m^2}\)
\item[\(A_V^R\)] effective valve restriction area, \(\mathrm{m^2}\)
\item[\(B_{\mathrm{lam}}\)] laminar-flow pressure ratio
\item[\(C\)] delayed-neutron precursor concentration
\item[\(C_d\)] discharge coefficient
\item[\(c_p\)] specific heat capacity, \(\mathrm{J\,kg^{-1}\,K^{-1}}\)
\item[\(D\)] diameter, \(\mathrm{m}\)
\item[\(e\)] control error
\item[\(h\)] specific enthalpy, \(\mathrm{J\,kg^{-1}}\)
\item[\(\Delta h_T\)] turbine specific enthalpy drop, \(\mathrm{J\,kg^{-1}}\)
\item[\(K\)] gain or resistance coefficient
\item[\(K_P,K_I\)] proportional and integral controller gains
\item[\(k_T\)] Stodola turbine constant
\item[\(L\)] length, \(\mathrm{m}\)
\item[\(L_1,L_2,L_3\)] subcooled, two-phase, and superheated SG region lengths, \(\mathrm{m}\)
\item[\(L_T\)] total steam-generator tube length, \(\mathrm{m}\)
\item[\(m\)] mass, \(\mathrm{kg}\)
\item[\(\dot{m}\)] mass flow rate, \(\mathrm{kg\,s^{-1}}\)
\item[\(M_T\)] turbine shaft torque, \(\mathrm{N\,m}\)
\item[\(N\)] number of helical tubes
\item[\(n_T\)] turbine polytropic exponent
\item[\(p\)] pressure, \(\mathrm{Pa}\)
\item[\(\Delta p\)] pressure difference, \(\mathrm{Pa}\)
\item[\(P\)] power, \(\mathrm{W}\)
\item[\(\dot{Q}\)] heat-transfer rate, \(\mathrm{W}\)
\item[\(s\)] specific entropy, \(\mathrm{J\,kg^{-1}\,K^{-1}}\)
\item[\(t\)] time, \(\mathrm{s}\)
\item[\(T\)] temperature, \({}^{\circ}\mathrm{C}\) or \(\mathrm{K}\)
\item[\(\bar{T}\)] average temperature
\item[\(u\)] specific internal energy, \(\mathrm{J\,kg^{-1}}\)
\item[\(V\)] volume, \(\mathrm{m^3}\)
\item[\(w\)] local flow speed, \(\mathrm{m\,s^{-1}}\)
\item[\(x_s\)] moving-boundary state vector
\item[\(z\)] axial coordinate along SG tube, \(\mathrm{m}\)
\end{description}

\vspace{0.4em}

\textit{Greek symbols}
\begin{description}[
    style=sameline,
    leftmargin=1.65cm,
    labelwidth=1.35cm,
    labelsep=0.3cm,
    itemsep=0pt,
    parsep=0pt,
    topsep=2pt
]
\item[\(\alpha_f,\alpha_c\)] fuel and coolant temperature reactivity coefficients, \({}^{\circ}\mathrm{C}^{-1}\)
\item[\(\alpha_i,\alpha_o\)] inner and outer heat-transfer coefficients, \(\mathrm{W\,m^{-2}\,K^{-1}}\)
\item[\(\alpha_V\)] void fraction at valve restriction
\item[\(\beta\)] effective delayed-neutron fraction
\item[\(\bar{\gamma}\)] average void fraction in the saturated region
\item[\(\eta\)] efficiency
\item[\(\eta_{\mathrm{isen}}\)] isentropic efficiency
\item[\(\eta_{\mathrm{mech}}\)] mechanical efficiency
\item[\(\eta_{\mathrm{SG}}\)] steam-generator heat-transfer efficiency
\item[\(\lambda\)] effective precursor decay constant, \(\mathrm{s^{-1}}\)
\item[\(\Lambda\)] prompt neutron lifetime, \(\mathrm{s}\)
\item[\(\nu\)] specific volume, \(\mathrm{m^3\,kg^{-1}}\)
\item[\(\rho\)] density, \(\mathrm{kg\,m^{-3}}\)
\item[\(\rho_{\mathrm{react}}\)] total core reactivity
\item[\(\rho_{\mathrm{ext}}\)] externally imposed control-rod reactivity
\item[\(\tau\)] time constant, \(\mathrm{s}\)
\item[\(\tau_f\)] fraction of reactor power deposited in fuel
\item[\(\phi\)] normalized neutron flux
\item[\(\Phi\)] energy flow rate, \(\mathrm{W}\)
\item[\(\psi_C\)] condenser liquid volume fraction
\item[\(\omega_T\)] turbine shaft angular speed, \(\mathrm{rad\,s^{-1}}\)
\end{description}

\vspace{0.4em}

\textit{Superscripts and subscripts}
\begin{description}[
    style=sameline,
    leftmargin=1.65cm,
    labelwidth=1.35cm,
    labelsep=0.3cm,
    itemsep=0pt,
    parsep=0pt,
    topsep=2pt
]
\item[\(0\)] initial value
\item[\(A,B\)] component ports
\item[\(C\)] condenser
\item[\(c,c1,c2\)] coolant and core coolant nodes
\item[\(CL,HL\)] cold leg and hot leg
\item[\(\mathrm{cmd}\)] commanded value
\item[\(\mathrm{ext}\)] externally imposed
\item[\(f\)] fuel
\item[\(\mathrm{fw}\)] feedwater or feedwater pump
\item[\(g,l\)] saturated vapor and saturated liquid
\item[\(i,o\)] inner and outer tube sides
\item[\(\mathrm{in},\mathrm{out}\)] inlet and outlet
\item[\(\mathrm{HP},\mathrm{LP}\)] high- and low-pressure turbine stages
\item[\(\mathrm{lin}\)] linear baseline model
\item[\(p\)] primary side
\item[\(R\)] controlled reservoir
\item[\(\mathrm{rated}\)] rated value
\item[\(\mathrm{ref}\)] reference value
\item[\(s\)] secondary side or steam
\item[\(\mathrm{sat}\)] saturated state
\item[\(\mathrm{sc}\)] subcooled region
\item[\(\mathrm{SG}\)] steam generator
\item[\(\mathrm{sh}\)] superheated region
\item[\(T\)] turbine
\item[\(V\)] valve
\item[\(w\)] tube wall
\end{description}

\end{multicols}

\end{minipage}}
\par
\endgroup

\vspace{1em}

\section{Introduction}\label{sec:introduction}

The transition to low-carbon power systems is increasing the need for firm, dispatchable generation technologies that can respond flexibly to changing demand. As power systems incorporate larger shares of variable renewable generation, flexible low-carbon resources are increasingly needed to support reliability while limiting emissions. Small modular reactors (SMRs), which are nuclear reactors built at a smaller scale than conventional plants, are being considered for this role~\cite{lher2024potential}. Future SMR deployment may therefore require operation beyond traditional baseload service, including load-following operation in response to changing grid or process demands~\cite{abusaleem2020issues}. Similar flexibility is also relevant for emerging energy-intensive applications such as hyperscale data centers, industrial energy systems, and maritime energy systems, where compact and reliable low-carbon generation may be valuable~\cite{you2025dynamic,cha2025potential,lirethinking,rahman2025multi,poudel2023design,senemmar2024navigating,badakhshan2025stochastic}. Across these applications, flexible operation depends not only on reactor-side response but also on the thermodynamic interaction between the reactor, steam generator (SG), and Rankine-cycle power-conversion system. Recent plant-wide SMR tests further demonstrate that flexible operation depends on coordinated reactor, SG, feedwater, and turbine-side pressure control during power maneuvers and trip events~\cite{dong2025testing}.

Integral pressurized water reactors (iPWRs) are an important class of water-cooled SMRs because they build on established pressurized water reactor technology while adopting a compact integral primary-system architecture~\cite{Arda2015Dynamic,Arda2016Nonlinear,Poudel2020Dynamic}. In these systems, major primary-side components are closely integrated, strengthening the dynamic coupling among reactor kinetics, primary-loop thermal hydraulics, SG heat transfer, and secondary-side power conversion. During load-following maneuvers, changes in turbine demand alter steam flow, secondary-side pressure, and turbine expansion conditions. These changes affect SG heat removal, primary-loop temperatures, and reactor reactivity feedback~\cite{zarei2020inherent,wang2026integrated}. Accurately capturing these interactions is therefore essential for evaluating transient response, control performance, thermodynamic efficiency, and operating margins~\cite{wu2025coordinated,zhang2024loadfollowing}. However, many existing dynamic studies either emphasize the nuclear steam supply system or the balance of plant (BOP), and studies that include both often represent the secondary cycle using simplified transfer functions, turbine-governor models, or prescribed steam-flow relationships~\cite{vajpayee2020dynamic}. Such simplifications can suppress the pressure--flow and enthalpy--flow couplings that govern off-design Rankine-cycle behavior.

Existing dynamic studies of SMRs span a wide range of model fidelity. On the reactor side, models ranging from reduced-order approximations to higher-fidelity formulations of reactor kinetics, primary-loop thermal hydraulics, and SG dynamics have been developed by Arda and Holbert~\cite{Arda2015Dynamic,Arda2016Nonlinear}, Poudel et al.~\cite{Poudel2020Dynamic}, Sabir et al.~\cite{Sabir2021LoadFrequency}, Sabir and Jiang~\cite{Sabir2022Comparing}, Ma et al.~\cite{Ma2019LoadFollowing}, Park et al.~\cite{park2023control}, Byun and Yim~\cite{byun2025variableTavg}, Wu et al.~\cite{Wu2024ComparisonStudy}, and Fakhraei et al.~\cite{Fakhraei2024DYSN}, often for load-following, frequency-control, or transient-control studies. In particular, the compact NuScale-type dynamic models developed by Arda and Holbert~\cite{Arda2015Dynamic,Arda2016Nonlinear} provide an important foundation for equation-based SMR modeling with point kinetics, lumped core thermal hydraulics, natural circulation, hot- and cold-leg transport, and moving-boundary helical-coil SG dynamics. Related extensions have used similar reactor--HCSG modeling structures for power-system and load-frequency-control studies, including electrically coupled multimodule SMR configurations~\cite{Poudel2020Dynamic,Sabir2021LoadFrequency,Sabir2022Comparing}. These studies have advanced the dynamic representation of SMR primary systems and steam generators. In many cases, however, the Rankine-cycle BOP is still represented by reduced-order turbine-governor models, selected feedwater or steam-side components, or control-oriented formulations rather than by a complete physics-based dynamic steam cycle.

Conversely, studies focused on the secondary side have provided detailed insight into SG and Rankine-cycle behavior, but they often consider selected components or operating functions rather than a fully coupled SMR--Rankine dynamic system. SG-focused studies include moving-boundary and modular models of helical-coil once-through SGs, as well as cogeneration-oriented SG operation and control formulations~\cite{Wu2023ThreeRegion,Bai2024OperationScheme,wei2026control,Kim2024HCSG}. Other studies emphasize broader secondary-cycle or hybrid-system operation, including nuclear hybrid energy-system architectures~\cite{Masotti2023NHES}, turbine and Rankine-cycle flexibility optimization~\cite{Vescovi2025Optimizing}, and steady-state comparisons of alternative SMR power cycles~\cite{Kissick2021Comparative}. These works provide valuable insight into SG dynamics, turbine-side flexibility, cogeneration, secondary-cycle behavior, and off-design operation. Nevertheless, the valve, turbine, condenser, and feedwater system are generally not coupled as a complete physics-based dynamic Rankine cycle to an equally detailed SMR primary-side model.

A limited number of recent studies have moved toward more integrated SMR and BOP simulation, including plant-level frameworks for cogeneration, coordinated control, power-system interaction, and flexible operation by Poudel and Gokaraju~\cite{poudel2021smr}, Wang et al.~\cite{wang2023coordinated}, Chen et al.~\cite{chen2024detailed}, Masotti et al.~\cite{Masotti2025Dynamic}, and Wang et al.~\cite{Wang2024DynamicSimulation}. These studies provide important advances toward plant-level integration. However, their emphasis is typically on hybrid-system operation, coordinated control, power-system interaction, turbine-speed response, or cogeneration strategy. Consequently, the SG and secondary cycle are often represented through quasi-static heating models, turbine-governor formulations, representative BOP architectures, three-lump SG models, or control-oriented component models rather than through an explicitly coupled moving-boundary SG and physics-based dynamic Rankine cycle. As a result, there remains a need for a plant-level load-following framework that combines a detailed equation-based SMR model capable of tracking the dynamic subcooled, two-phase, and superheated SG regions with a physics-based secondary Rankine cycle, while also quantifying the consequences of simplified secondary-side representations.

\begin{figure*}[tbp!]
  \centering
  \includegraphics[width=\linewidth]{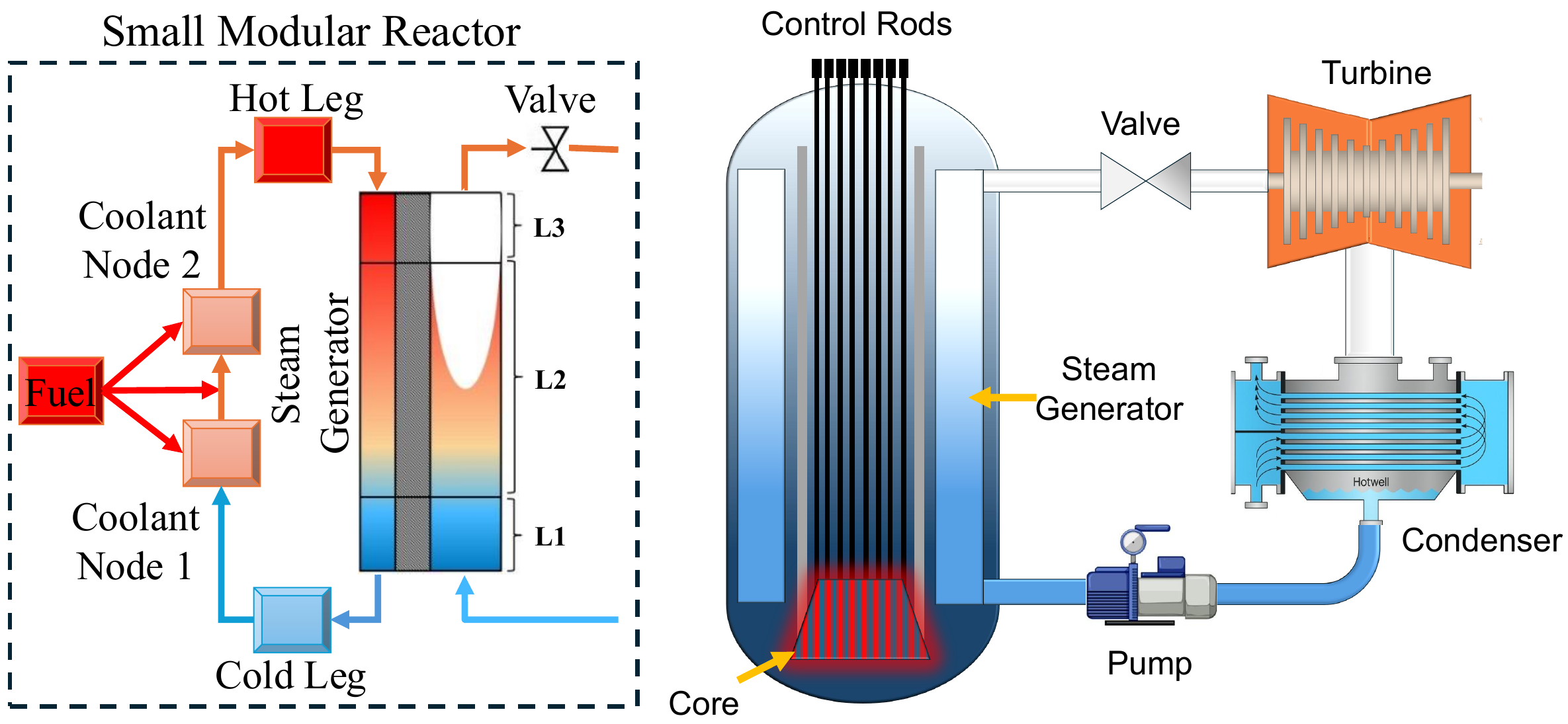} 
  \caption{Integrated dynamic model of the SMR and Rankine steam cycle. The equation-based SMR primary-system model includes point kinetics, fuel and coolant temperature feedback, lumped core thermal hydraulics, natural circulation, hot- and cold-leg transport, and a moving-boundary helical-coil once-through SG with subcooled ($L_1$), two-phase ($L_2$), and superheated ($L_3$) regions. The secondary Rankine cycle includes the steam throttle valve, turbine, condenser, and feedwater pump, which are thermodynamically coupled to the SG to capture dynamic back-pressure and off-design turbine behavior during load-following transients.}
  \label{fig:Overview_Figure_1}
\end{figure*}

This study addresses this gap by developing a thermodynamically coupled dynamic framework for an iPWR-based SMR and Rankine steam cycle, as illustrated in Fig.~\ref{fig:Overview_Figure_1}. The framework couples an equation-based model of the NuScale-type iPWR, including reactor kinetics, primary-loop thermal hydraulics, natural circulation, hot- and cold-leg transport, and a moving-boundary helical-coil once-through SG, with a physics-based secondary Rankine cycle implemented in Simscape. The secondary cycle includes the steam throttle valve, turbine, condenser, and feedwater pump. Unlike conventional signal-flow BOP models, the proposed framework enforces mass and energy conservation across the coupled domains and allows steam flow, back-pressure, turbine enthalpy drop, and condenser inventory to evolve from the thermodynamic state of the system.

The main contributions of this work are summarized as follows:
\begin{enumerate}
    \item A hybrid SMR--Rankine-cycle dynamic modeling framework is developed by coupling an equation-based reactor and moving-boundary helical-coil steam-generator model with a physics-based secondary Rankine cycle.

    \item The proposed framework replaces simplified steam-flow and turbine-governor representations with thermodynamic valve, turbine, condenser, and feedwater-pump components that capture pressure--flow coupling, load-dependent turbine enthalpy drop, condenser inventory dynamics, and feedwater-system response.

    \item A decentralized load-following control architecture is implemented to coordinate the steam throttle valve, feedwater pump, and control rods for simultaneous turbine mechanical-power tracking, SG pressure regulation, and primary coolant temperature control.

    \item The physics-based framework is compared with a conventional linear transfer-function secondary-cycle baseline under the same SMR formulation and control objectives, allowing the thermodynamic gap introduced by simplified secondary-cycle assumptions to be isolated and quantified.

    \item Simulation results show that models with nearly identical mechanical-power tracking can predict different steam-flow requirements, reactor thermal power levels, and steam-generator moving-boundary behavior, demonstrating that external power response alone is insufficient for assessing SMR load-following performance and SG operating margins.
\end{enumerate}

The remainder of this paper is organized as follows. Section~\ref{Sec:SMR-model} presents the mathematical formulation of the SMR primary system and moving-boundary helical-coil steam generator. Section~\ref{Sec:steam-cycle} describes the physics-based modeling of the secondary Rankine-cycle components, including the steam throttle valve, turbine, condenser, and feedwater pump. Section~\ref{Sec:integrated-control} presents the thermodynamic coupling interface and the decentralized load-following control architecture. Section~\ref{Sec:results} analyzes the dynamic response under different control configurations and compares the proposed physics-based framework with a simplified linear transfer-function steam-cycle model. Finally, Section~\ref{Sec:conclusion} summarizes the main findings and discusses future extensions.


\section{Small modular reactor modeling}\label{Sec:SMR-model}

The SMR primary system and moving-boundary SG are modeled using a compact
equation-based formulation that builds on the NuScale-type dynamic modeling framework
of Arda and Holbert~\cite{Arda2015Dynamic,Arda2016Nonlinear}. Parameterization and
coupling modifications are introduced here to enable integration with the physics-based
secondary Rankine cycle. Water and steam thermodynamic properties required in the
equation-based SMR and SG model were evaluated using the XSteam MATLAB function,
which implements the IAPWS-IF97 formulation for water and steam properties
\cite{HolmgrenXSteam,IAPWS1997}. XSteam was used to obtain temperatures, densities,
enthalpies, saturation properties, and property derivatives from local
pressure--temperature or pressure--enthalpy states.


\subsection{Reactor model}\label{Sec:reactor-model}

The reactor model represents the NuScale-type iPWR using a compact equation-based formulation that includes point-kinetics neutronics, temperature reactivity feedback, lumped core thermal hydraulics, buoyancy-driven natural circulation, and hot- and cold-leg transport. This level of fidelity was selected to preserve the dominant reactor and primary-loop dynamics relevant to load-following operation while maintaining computational tractability for plant-level coupling with the secondary Rankine cycle.

\subsubsection{Core neutronics and reactivity feedback}\label{Sec:core-neutronics}

Core neutronics were described using a lumped point-kinetics formulation with one effective delayed-neutron precursor group obtained by aggregating the six precursor groups. The neutron population was represented by the normalized average neutron flux, \(\phi\), such that \(\phi=1\) corresponds to rated power. The effective precursor concentration is denoted by \(C\). The point-kinetics equations are:
\begin{align}
\frac{\mathrm{d}\phi}{\mathrm{d}t} &=
\left(\frac{\rho_{\mathrm{react}}-\beta}{\Lambda}\right)\phi+\lambda C, \\
\frac{\mathrm{d}C}{\mathrm{d}t} &=
\frac{\beta}{\Lambda}\phi-\lambda C,
\end{align}
where \(\rho_{\mathrm{react}}\) is the total core reactivity, \(\beta\) is the effective delayed-neutron fraction, \(\Lambda\) is the prompt neutron lifetime, and \(\lambda\) is the effective precursor decay constant. Reactor thermal power was then obtained from the normalized flux as:

\begin{equation}
P_{\mathrm{th}}=\phi\times P_{th}^{\mathrm{rated}},
\end{equation}
where \(P_{th}^{\mathrm{rated}}\) is the rated reactor thermal power.

The total reactivity combines externally imposed rod reactivity with temperature feedback from the fuel and primary coolant. It is written as:
\begin{equation}
\rho_{\mathrm{react}}
=
\rho_{\mathrm{ext}}
+
\alpha_f \Delta T_f
+
\alpha_c
\frac{\Delta T_{c1}+\Delta T_{c2}}{2},
\end{equation}
where \(\rho_{\mathrm{ext}}\) is the externally imposed control-rod reactivity, \(\alpha_f\) and \(\alpha_c\) are the fuel and coolant temperature reactivity coefficients, respectively, and \(\Delta T_f\), \(\Delta T_{c1}\), and \(\Delta T_{c2}\) are deviations from the corresponding nominal temperatures. The two coolant temperatures \(T_{c1}\) and \(T_{c2}\) represent the lower and upper core coolant nodes, so their average provides the moderator-temperature feedback signal used in the reactivity balance. The numerical values used for these parameters are listed in Table~\ref{tab:neutronics_params}.

\begin{table}[thb!]
\caption{Neutron-kinetics and reactivity-feedback parameters.}
\label{tab:neutronics_params}
\centering
\small
\renewcommand{\arraystretch}{1.15}
\begin{tabular}{lll}
\toprule
\textbf{Parameter} & \textbf{Description} & \textbf{Value} \\
\midrule
\(\alpha_f\) & Fuel temperature reactivity coefficient & \(-2.16\times10^{-5}~^\circ\mathrm{C}^{-1}\) \\
\(\alpha_c\) & Coolant/moderator temperature reactivity coefficient & \(-1.8\times10^{-4}~^\circ\mathrm{C}^{-1}\) \\
\(\Lambda\) & Prompt neutron generation time & \(2.0\times10^{-5}~\mathrm{s}\) \\
\(\beta\) & Effective delayed-neutron fraction & \(0.007\) \\
\(\lambda\) & Effective precursor decay constant & \(0.1~\mathrm{s}^{-1}\) \\
\bottomrule
\end{tabular}
\end{table}

\subsubsection{Core thermal hydraulics and natural circulation}\label{Sec:core-thermal}

The core thermal hydraulics were modeled using a reduced lumped-parameter form of Mann's model \cite{Kerlin1976Robinson}. A single fuel node was thermally coupled to two primary-coolant nodes, providing a compact representation of heat generation, fuel-to-coolant heat transfer and axial coolant heating through the core. The main reactor-power, core thermal-hydraulic and primary-loop transport parameters used in these equations are listed in Table~\ref{tab:core_thermal_params}. The governing energy balances are:
\begin{align}
\frac{\mathrm{d}T_f}{\mathrm{d}t}
&=
\frac{\tau_f P_{\mathrm{th}}
+
h_{fc}A_{fc}\left(T_{c1}-T_f\right)}
{m_f c_{pf}},
\\[4pt]
\frac{\mathrm{d}T_{c1}}{\mathrm{d}t}
&=
\frac{(1-\tau_f)P_{\mathrm{th}}
+
h_{fc}A_{fc}\left(T_f-T_{c1}\right)}
{m_c c_{pc}}
+
\frac{2\dot{m}_p\left(h_{CL}-h_{c1}\right)}
{m_c c_{pc}},
\\[4pt]
\frac{\mathrm{d}T_{c2}}{\mathrm{d}t}
&=
\frac{(1-\tau_f)P_{\mathrm{th}}
+
h_{fc}A_{fc}\left(T_f-T_{c1}\right)}
{m_c c_{pc}}
+
\frac{2\dot{m}_p\left(h_{c1}-h_{c2}\right)}
{m_c c_{pc}} .
\end{align}
Here, \(T_f\) is the lumped fuel temperature; \(T_{c1}\) and \(T_{c2}\) are the two primary-coolant node temperatures; \(m_f\) and \(m_c\) are the effective fuel and coolant masses; \(c_{pf}\) and \(c_{pc}\) are the corresponding heat capacities; \(h_{fc}A_{fc}\) is the fuel-to-coolant heat-transfer conductance; and \(\tau_f\) is the fraction of reactor power deposited directly in the fuel. The primary-loop mass flow rate is denoted by \(\dot{m}_p\), while \(h_{CL}\), \(h_{c1}\), and \(h_{c2}\) are the specific enthalpies associated with the cold leg and the two coolant nodes. The coolant thermophysical properties were evaluated from water-property functions at the primary-loop pressure and local thermal state.

Because the NuScale-type iPWR relies on buoyancy-driven natural circulation, the primary flow rate was not imposed by a pump model. Instead, the steady primary mass flow was scaled with reactor power using the cube-root relation commonly adopted for natural-circulation SMR models \cite{Arda2015Dynamic}. With the neutron flux normalized to rated power, the primary mass flow rate was written as:
\begin{equation}
\dot{m}_p
=
\dot{m}_{p}^{\mathrm{rated}}\phi^{1/3},
\end{equation}
where \(\dot{m}_{p}^{\mathrm{rated}}\) is the rated primary mass flow rate.

\subsubsection{Hot-leg and cold-leg transport}\label{Sec:loop-transport}

Transport delays in the primary loop were represented using first-order lag models for the hot and cold legs. The hot-leg temperature, \(T_{HL}\), relaxes toward the core outlet temperature \(T_{c2}\), whereas the cold-leg temperature, \(T_{CL}\), relaxes toward the primary-side outlet temperature of the steam generator, \(T_{p6}\). The corresponding transport equations are:
\begin{align}
\frac{\mathrm{d}T_{HL}}{\mathrm{d}t}
&=
\frac{T_{c2}-T_{HL}}{\tau_{HL}},
&
\tau_{HL}
&=
\frac{m_{HL}}{\dot{m}_p},
\\[4pt]
\frac{\mathrm{d}T_{CL}}{\mathrm{d}t}
&=
\frac{T_{p6}-T_{CL}}{\tau_{CL}},
&
\tau_{CL}
&=
\frac{m_{CL}}{\dot{m}_p}.
\end{align}
Here, \(m_{HL}\) and \(m_{CL}\) are the effective coolant mass inventories of the hot and cold legs, respectively. These inventories were computed from the corresponding coolant volumes in Table~\ref{tab:core_thermal_params} and the primary-coolant density at the local thermal state. These first-order transport elements introduce the thermal inertia and delay associated with coolant circulation between the reactor core and the SG, which are important for reproducing the timing of primary-temperature feedback during load-following transients.

\begin{table}[thbp!]
\caption{Reactor-power, core thermal-hydraulic, and primary-loop transport parameters.}
\label{tab:core_thermal_params}
\centering
\small
\renewcommand{\arraystretch}{1.15}
\begin{tabular}{lll}
\toprule
\textbf{Parameter} & \textbf{Description} & \textbf{Value} \\
\midrule
\(P_{\mathrm{th}}^{\mathrm{rated}}\) & Rated reactor thermal power & \(160~\mathrm{MW}\) \\
\(\tau_f\) & Fraction of reactor power deposited in fuel & \(0.97\) \\
\(\dot{m}_{p}^{\mathrm{rated}}\) & Rated primary mass flow rate & \(587~\mathrm{kg\,s^{-1}}\) \\
\(p_p\) & Primary-loop pressure & \(12.76~\mathrm{MPa}\) \\
\(m_f\) & Effective fuel mass & \(11252~\mathrm{kg}\) \\
\(c_{pf}\) & Fuel specific heat capacity & \(467~\mathrm{J\,kg^{-1}\,K^{-1}}\) \\
\(h_{fc}\) & Fuel-to-coolant heat-transfer coefficient & \(1135~\mathrm{W\,m^{-2}\,K^{-1}}\) \\
\(A_{fc}\) & Fuel-to-coolant heat-transfer area & \(583~\mathrm{m^2}\) \\
\(V_{\mathrm{core}}\) & Core coolant volume & \(2.5202~\mathrm{m^3}\) \\
\(V_{HL}\) & Hot-leg coolant volume & \(17.9812~\mathrm{m^3}\) \\
\(V_{CL}\) & Cold-leg coolant volume & \(16.3671~\mathrm{m^3}\) \\
\bottomrule
\end{tabular}
\end{table}

\subsection{Moving-boundary steam generator model}\label{Sec:SG-model}

\subsubsection{Model structure and assumptions}\label{Sec:SG-assumptions}

The helical-coil once-through SG was represented using a counterflow moving-boundary formulation. The primary coolant flows downward on the shell side as a single-phase liquid, while the secondary fluid flows upward through the helical tubes and is heated from subcooled liquid to saturated two-phase flow and then to dry or superheated steam. The secondary side was divided into three moving regions: a subcooled region of length \(L_1\), a saturated boiling region of length \(L_2\), and a superheated region of length \(L_3\). The total tube length was fixed, so that
\begin{equation}
L_1+L_2+L_3=L_T .
\end{equation}

The moving-boundary formulation provides a compact representation of the phase-change dynamics that govern SG response during load-following transients. Each secondary-side region was treated as a control volume, and the primary coolant and tube metal were discretized consistently with the same three-region structure. One representative temperature node was used for the tube metal in each region, denoted by \(T_{w1}\), \(T_{w2}\), and \(T_{w3}\), and one representative primary-side temperature node was used for each corresponding shell-side region, denoted by \(T_{p1}\), \(T_{p2}\), and \(T_{p3}\). The model assumes one-dimensional flow, uniform secondary-side pressure, and a single-tube equivalent representation scaled by the total number of helical tubes, \(N\). Table~\ref{tab:sg_params} lists the SG geometric, heat-transfer, and tube-metal parameters used in the model, with the geometric parameters taken from the NuScale Standard Plant Design Certification Application \citep{nuscale2020dca}.

\begin{table}[thb!]
\caption{Steam generator geometry and heat-transfer parameters.}
\label{tab:sg_params}
\centering
\small
\renewcommand{\arraystretch}{1.15}
\begin{tabular}{lll}
\toprule
\textbf{Parameter} & \textbf{Description} & \textbf{Value} \\
\midrule
\(N\) & Number of helical tubes & \(1380~(690\times2)\) \\
\(D_o\) & Tube outer diameter & \(0.0159~\mathrm{m}\) \\
\(D_i\) & Tube inner diameter & \(0.01336~\mathrm{m}\) \\
\(L_T\) & Total tube length & \(24.2~\mathrm{m}\) \\
\(A_w\) & Tube-wall cross-sectional area & \(\frac{\pi}{4}(D_o^2-D_i^2)\) \\
\(A_p\) & Primary-side flow area & \(0.7266~\mathrm{m^2}\) \\
\(A_s\) & Secondary-side flow area & \(0.1935~\mathrm{m^2}\) \\
\(\alpha_o\) & Primary-side heat-transfer coefficient & \(19093~\mathrm{W\,m^{-2}\,K^{-1}}\) \\
\(\alpha_i\) & Secondary-side heat-transfer coefficient & \(2697~\mathrm{W\,m^{-2}\,K^{-1}}\) \\
\(\eta_{SG}\) & steam generator heat-transfer efficiency & \(0.9946\) \\
\(\rho_w\) & Tube-wall density & \(8192~\mathrm{kg\,m^{-3}}\) \\
\(c_w\) & Tube-wall specific heat capacity & \(463~\mathrm{J\,kg^{-1}\,K^{-1}}\) \\
\(\bar{\gamma}\) & Average void fraction in saturated region & \(0.251913\) \\
\bottomrule
\end{tabular}
\end{table}

\subsubsection{Secondary-side moving-boundary equations}\label{Sec:SG-secondary}

The secondary-side dynamics were obtained by applying one-dimensional mass and energy conservation to the tube-side fluid. The governing conservation equations are:
\begin{align}
\frac{\partial \left(A_s \rho_s\right)}{\partial t}
+
\frac{\partial \dot{m}_s}{\partial z}
&=0,
\\[3pt]
\frac{\partial \left[A_s\left(\rho_s h_s-p_s\right)\right]}{\partial t}
+
\frac{\partial \left(\dot{m}_s h_s\right)}{\partial z}
&=
\pi D_i \alpha_i\left(T_w-\bar{T}_s\right),
\end{align}
where \(z\) is the axial coordinate along the tube, \(A_s\) is the secondary-side flow area, \(\rho_s\) is the secondary-fluid density, \(h_s\) is the specific enthalpy, \(p_s\) is the secondary-side pressure, and \(\dot{m}_s\) is the secondary mass flow rate. The right-hand side of the energy equation represents heat transfer from the tube wall to the secondary fluid, where \(D_i\) is the tube inner diameter, \(\alpha_i\) is the inner heat-transfer coefficient, \(T_w\) is the wall temperature, and \(\bar{T}_s\) is the regional average secondary-fluid temperature.

The conservation equations were integrated over the subcooled, saturated, and superheated regions using Leibniz's rule. This produces a set of ordinary differential equations for the moving boundaries and thermodynamic states. The secondary-side state vector was defined as:
\begin{equation}
x_s =
\begin{bmatrix}
L_1 & L_2 & p_s & h_{sh}^{out}
\end{bmatrix}^{T},
\end{equation}
where \(h_{sh}^{out}\) is the outlet enthalpy of the superheated region. The resulting moving-boundary model can be written compactly as:
\begin{equation}
\mathbf{A}(x_s)\dot{x}_s=\mathbf{b}(u_s),
\end{equation}
or equivalently,
\begin{equation}
\begin{bmatrix}
a_{1,1} & a_{1,2} & a_{1,3} & a_{1,4} \\
a_{2,1} & 0       & a_{2,3} & 0       \\
a_{3,1} & a_{3,2} & a_{3,3} & a_{3,4} \\
a_{4,1} & a_{4,2} & a_{4,3} & a_{4,4}
\end{bmatrix}
\begin{bmatrix}
\dot{L}_1 \\
\dot{L}_2 \\
\dot{p}_s \\
\dot{h}_{sh}^{out}
\end{bmatrix}
=
\begin{bmatrix}
b_1 \\
b_2 \\
b_3 \\
b_4
\end{bmatrix}.
\end{equation}
The coefficient matrix \(\mathbf{A}\) contains the thermodynamic derivatives and geometric terms associated with the three secondary-side regions, while \(\mathbf{b}\) contains the inlet and outlet mass-flow terms and the heat transferred from the tube wall. The inlet mass flow rate, \(\dot{m}_s^{in}\), is supplied by the feedwater pump model, and the outlet steam flow rate, \(\dot{m}_s^{out}\), is determined by the downstream steam-valve and turbine model. The inlet enthalpy \(h_s^{in}\) defines the feedwater thermodynamic state at the SG inlet. Saturated liquid and vapor properties are evaluated at the instantaneous secondary pressure \(p_s\), and the subcooled and superheated densities are updated from the corresponding pressure--enthalpy states. The coefficient definitions used in the secondary-side moving-boundary model are given in Table~\ref{tab:matrix_coeffs}. These coefficients follow the moving-boundary formulation in \cite{Arda2016Nonlinear}, with the heat-transfer terms scaled by the total number of helical tubes, \(N\).

\begin{table}[!thb]
\caption{Matrix coefficients for secondary side dynamics.}
\label{tab:matrix_coeffs}
\centering
\renewcommand{\arraystretch}{1.12}
\begin{tabular}{@{}ll@{}}
\toprule
\textbf{Coefficient} & \textbf{Expression} \\ 
\midrule
\multicolumn{2}{@{}l}{\textit{Row 1: Mass Balance}} \\
$a_{1,1}$ & $A_s(\rho_{\mathrm{sc}} - \rho_{\mathrm{sh}})$ \\
$a_{1,2}$ & $A_s\big[(1 - \bar{\gamma})\rho_l + \bar{\gamma}\rho_g - \rho_{\mathrm{sh}}\big]$ \\
$a_{1,3}$ & $A_s\Big[ L_1 \xi_{sc} + L_2 \xi_{sat2} + L_3 \xi_{sh} \Big]$ \\
$a_{1,4}$ & $\frac{1}{2} A_s L_3 \frac{\partial \rho_{\mathrm{sh}}}{\partial h}$ \\
$b_1$ & $\dot{m}_{s}^{in} - \dot{m}_{s}^{out}$ \\
\addlinespace
\multicolumn{2}{@{}l}{\textit{Row 2: Subcooled Energy Balance}} \\
$a_{2,1}$ & $A_s \rho_{sc} (h_{sc}^{in} - h_l)$ \\
$a_{2,3}$ & $A_s L_1 \left[ \frac{1}{2} \rho_{sc} \frac{\partial h_l}{\partial p_s} + (h_{sc}^{in} - h_l)\xi_{sc} - 1 \right]$ \\
$b_2$ & $\dot{m}_{s}^{in} (h_{sc}^{in} - h_l) + N \pi D_i \alpha_i L_1 (T_{w1} - T_{sc})$ \\
\addlinespace
\multicolumn{2}{@{}l}{\textit{Row 3: Saturated Energy Balance}} \\
$a_{3,1}$ & $A_s (\rho_{sc} h_l - \rho_{sh} h_g)$ \\
$a_{3,2}$ & $A_s \left[ (1 - \bar{\gamma})\rho_l h_l + \bar{\gamma}\rho_g h_g - \rho_{sh} h_g \right]$ \\
$a_{3,3}$ & $A_s \Big[ h_l L_1 \xi_{sc} + L_2 \xi_{sat1} + h_g L_3 \xi_{sh} \Big]$ \\
$a_{3,4}$ & $A_s L_3 \frac{1}{2} h_g \frac{\partial \rho_{sh}}{\partial h}$ \\
$b_3$ & $\dot{m}_{s}^{in} h_l - \dot{m}_{s}^{out} h_g + N \pi D_i \alpha_i L_2 (T_{w2} - T_{sat})$ \\
\addlinespace
\multicolumn{2}{@{}l}{\textit{Row 4: Superheated Energy Balance}} \\
$a_{4,1}$ & $A_s \rho_{sh} (h_g - h_{sh}^{out})$ \\
$a_{4,2}$ & $A_s \rho_{sh} (h_g - h_{sh}^{out})$ \\
$a_{4,3}$ & $A_s L_3 \left[ \frac{1}{2} \rho_{sh} \frac{\partial h_g}{\partial p_s} + (h_{sh}^{out} - h_g) \xi_{sh} - 1 \right]$ \\
$a_{4,4}$ & $\frac{1}{2} A_s L_3 \left[ (h_{sh}^{out} - h_g) \frac{\partial \rho_{sh}}{\partial h} + \rho_{sh} \right]$ \\
$b_4$ & $\dot{m}_{s}^{out} (h_g - h_{sh}^{out}) + N \pi D_i \alpha_i L_3 (T_{w3} - T_{sh})$ \\
\addlinespace
\multicolumn{2}{@{}l}{\textit{Auxiliary Variables}} \\
$\xi_{sc}$ & $\frac{\partial \rho_{\mathrm{sc}}}{\partial p_s} + \frac{1}{2}\frac{\partial \rho_{\mathrm{sc}}}{\partial h}\frac{\partial h_l}{\partial p_s}$ \\
$\xi_{sh}$ & $\frac{\partial \rho_{\mathrm{sh}}}{\partial p_s} + \frac{1}{2}\frac{\partial \rho_{\mathrm{sh}}}{\partial h}\frac{\partial h_g}{\partial p_s}$ \\
$\xi_{sat1}$ & $\bar{\gamma} \frac{\partial (\rho_g h_g)}{\partial p_s} + (1 - \bar{\gamma}) \frac{\partial (\rho_l h_l)}{\partial p_s} - 1$ \\
$\xi_{sat2}$ & $\bar{\gamma} \frac{\partial \rho_g}{\partial p_s} + (1 - \bar{\gamma}) \frac{\partial \rho_l}{\partial p_s}$ \\
\bottomrule
\end{tabular}
\end{table}

In these expressions, \(\rho_{sc}\) and \(\rho_{sh}\) are the subcooled and superheated densities, respectively; \(\rho_l\) and \(\rho_g\) are the saturated liquid and vapor densities; \(h_l\) and \(h_g\) are the saturated liquid and vapor enthalpies; and \(\bar{\gamma}\) is the average void fraction in the saturated region. The auxiliary variables \(\xi_{sc}\), \(\xi_{sh}\), \(\xi_{sat1}\) and \(\xi_{sat2}\) collect the pressure and enthalpy derivatives of the water/steam properties that arise when the conservation equations are integrated over moving control volumes.

\subsubsection{Tube-metal energy balance}\label{Sec:SG-wall}

The tube metal was modeled as the thermal interface between the primary and secondary fluids. For each moving region, the wall temperature evolves according to a lumped radial energy balance that includes heat received from the primary coolant, heat transferred to the secondary fluid, and the apparent energy transport associated with motion of the regional boundaries. The general wall-energy balance is:
\begin{equation}
\rho_w c_w A_w \frac{\mathrm{d}T_w}{\mathrm{d}t}
=
\eta_{SG}\pi D_o\alpha_o\left(\bar{T}_p-T_w\right)
+
\pi D_i\alpha_i\left(\bar{T}_s-T_w\right),
\end{equation}
where \(\rho_w\), \(c_w\), and \(A_w\) are the wall density, specific heat capacity, and cross-sectional area, respectively. The outer tube diameter is \(D_o\), the primary-side heat-transfer coefficient is \(\alpha_o\), and \(\eta_{SG}\) accounts for SG heat-transfer efficiency or heat losses. The regional tube-metal balances are:
\begin{align}
\dot{T}_{w1}
&=
\frac{
N\pi\left[
\eta_{SG}D_o\alpha_o\left(T_{p1}-T_{w1}\right)
+
D_i\alpha_i\left(T_{sc}-T_{w1}\right)
\right]
}
{\rho_w c_w A_w}
-
\frac{T_{w1}-T_{w2}}{L_1}\dot{L}_1,
\\[5pt]
\dot{T}_{w2}
&=
\frac{
N\pi\left[
\eta_{SG}D_o\alpha_o\left(T_{p2}-T_{w2}\right)
+
D_i\alpha_i\left(T_{sat}-T_{w2}\right)
\right]
}
{\rho_w c_w A_w},
\\[5pt]
\dot{T}_{w3}
&=
\frac{
N\pi\left[
\eta_{SG}D_o\alpha_o\left(T_{p3}-T_{w3}\right)
+
D_i\alpha_i\left(T_{sh}-T_{w3}\right)
\right]
}
{\rho_w c_w A_w}
-
\frac{T_{w2}-T_{w3}}{L_3}
\left(\dot{L}_1+\dot{L}_2\right).
\end{align}
Here, \(T_{sc}\), \(T_{sat}\), and \(T_{sh}\) are representative secondary-fluid temperatures in the subcooled, saturated, and superheated regions, respectively. The terms proportional to \(\dot{L}_1\) and \(\dot{L}_1+\dot{L}_2\) account for the redistribution of tube-metal thermal inventory as the phase boundaries move. Since \(L_T\) is fixed, \(\dot{L}_3=-(\dot{L}_1+\dot{L}_2)\).

\subsubsection{Primary-side shell energy balance}\label{Sec:SG-primary}

The primary coolant was treated as a single-phase liquid flowing downward on the shell side of the SG. Mass storage in the primary-side SG nodes was neglected, and each regional node was represented by a lumped energy balance. The general form is:
\begin{equation}
\rho_p c_p A_p
\frac{\mathrm{d}\bar{T}_p}{\mathrm{d}t}
=
\pi D_o\alpha_o\left(T_w-\bar{T}_p\right)
+
\dot{m}_p
\left(h_p^{in}-h_p^{out}\right),
\end{equation}
where \(\rho_p\), \(c_p\), and \(A_p\) are the density, heat capacity, and effective flow area of the primary coolant, respectively, and \(\dot{m}_p\) is the primary mass flow rate. Nodal temperatures were obtained from the local pressure and average specific enthalpy using the water/steam property formulation:
\begin{equation}
\bar{T}_p
=
T\left(
p_p,
\frac{h_p^{in}+h_p^{out}}{2}
\right),
\end{equation}
where \(T(p,h)\) denotes the temperature returned by the water/steam property formulation for a given pressure and specific enthalpy.

Because the primary flow is countercurrent to the secondary flow, the primary coolant enters the SG at the hot-leg side, corresponding to the superheated secondary region, and exits near the subcooled secondary region. The regional primary-side balances are:
\begin{align}
\dot{T}_{p1}
&=
\frac{
N\pi D_o\alpha_o\left(T_{w1}-T_{p1}\right)
}
{\rho_{p1} c_{p1} A_p}
+
\frac{
\dot{m}_p\left(h_{p1}^{in}-h_{p1}^{out}\right)
}
{\rho_{p1} c_{p1} A_p L_1}
-
\frac{T_{p1}-T_{p2}}{L_1}\dot{L}_1,
\\[5pt]
\dot{T}_{p2}
&=
\frac{
N\pi D_o\alpha_o\left(T_{w2}-T_{p2}\right)
}
{\rho_{p2} c_{p2} A_p}
+
\frac{
\dot{m}_p\left(h_{p2}^{in}-h_{p2}^{out}\right)
}
{\rho_{p2} c_{p2} A_p L_2},
\\[5pt]
\dot{T}_{p3}
&=
\frac{
N\pi D_o\alpha_o\left(T_{w3}-T_{p3}\right)
}
{\rho_{p3} c_{p3} A_p}
+
\frac{
\dot{m}_p\left(h_{p3}^{in}-h_{p3}^{out}\right)
}
{\rho_{p3} c_{p3} A_p L_3}
-
\frac{T_{p2}-T_{p3}}{L_3}
\left(\dot{L}_1+\dot{L}_2\right).
\end{align}
The inlet enthalpy to the third primary-side node is the hot-leg enthalpy, \(h_{p3}^{in}=h_{HL}\). The outlet of the first primary-side node provides the cold-side thermal state returned to the primary loop. As in the wall model, the moving-boundary correction terms account for the changing effective control-volume lengths associated with the secondary-side phase-boundary motion.

\section{Physics-based Rankine-cycle modeling}\label{Sec:steam-cycle}

The secondary Rankine cycle is modeled using two-phase Simscape Fluids components for the steam throttle valve, turbine, condenser, and feedwater pump. These components are connected through acausal two-phase fluid ports, allowing pressure, mass flow rate, enthalpy, phase state, and energy exchange to be solved simultaneously from the conservation equations of the network. This formulation differs from causal transfer-function or signal-flow representations in which steam flow, turbine power, feedwater response, or condenser pressure are prescribed through fitted gains, time constants, algebraic maps, or fixed boundary conditions. In the present model, the valve flow depends on the instantaneous restriction area, pressure difference, and fluid state; the turbine power depends on the evolving pressure ratio and enthalpy drop; the condenser retains finite mass and energy storage; and the pump imposes feedwater flow while preserving the pressure and enthalpy changes required by the surrounding two-phase network. As a result, pressure--flow coupling, dynamic back-pressure, condenser inventory, throttling effects, and load-dependent turbine specific work emerge from the coupled thermodynamic solution rather than being imposed externally. This distinction is important for load-following analysis because the secondary cycle defines the pressure, flow, and enthalpy boundary conditions imposed on the moving-boundary SG.

\subsection{Steam throttle valve model}\label{Sec:valve-model}

The steam throttle valve upstream of the turbine was represented as an adiabatic two-phase local restriction using the Valve (Two-Phase) block in Simscape Fluids \cite{SimscapeFluidsR2023a}. The manipulated variable is the effective restriction area, \(A_V^R\), which is adjusted by the turbine power controller. The valve restriction-area limits, port area, discharge coefficient and laminar--turbulent blending parameter used in the model are listed in Table~\ref{tab:valve_params}.

\begin{table}[thbp!]
\caption{Steam-throttle-valve model parameters.}
\label{tab:valve_params}
\centering
\small
\renewcommand{\arraystretch}{1.15}
\begin{tabular}{lll}
\toprule
\textbf{Parameter} & \textbf{Description} & \textbf{Value} \\
\midrule
\(A_{V,\min}^{R}\) & Minimum restriction area & \(1.0\times10^{-5}~\mathrm{m^2}\) \\
\(A_{V,\max}^{R}\) & Maximum restriction area & \(5.0\times10^{-2}~\mathrm{m^2}\) \\
\(A_V^{port}\) & Cross-sectional area at valve ports A and B & \(8.0\times10^{-2}~\mathrm{m^2}\) \\
\(C_d\) & Discharge coefficient & \(0.64\) \\
\(B_{lam}\) & Laminar-flow pressure ratio & \(0.999\) \\
\bottomrule
\end{tabular}
\end{table}

The valve has no internal mass storage, so conservation of mass requires:
\begin{equation}
\dot{m}_{V}^{in}+\dot{m}_{V}^{out}=0 ,
\end{equation}
where \(\dot{m}_{V}^{in}\) and \(\dot{m}_{V}^{out}\) are the mass flow rates entering and leaving the valve ports, respectively. Because the restriction is treated as adiabatic, the net energy flux through the valve is zero.
\begin{equation}
\Phi_{V}^{in}+\Phi_{V}^{out}=0 .
\end{equation}

The valve therefore conserves total specific enthalpy across the inlet, restriction, and outlet states.
\begin{equation}
h_{V}^{\mathrm{tot,in}} = h_{V}^{\mathrm{tot,out}}, \qquad h_V^{\mathrm{tot}} = u_V+p_V\nu_V+\frac{w_V^2}{2},
\end{equation}
where \(u_V\) is the specific internal energy, \(p_V\) is the pressure, \(\nu_V\) is the specific volume, and \(w_V\) is the local flow speed. This expression separates the thermodynamic flow work, \(p_V\nu_V\), from the kinetic-energy contribution and avoids using the same symbol for specific volume and velocity.

The mass flow through the restriction was computed from a pressure-dependent momentum relation of the form
\begin{equation}
\dot{m}_{V}^{in}
=
C_d A_V^R \Delta p_V
\sqrt{
\frac{2}
{K_V(\Delta p_V)|\Delta p_V|\nu_V^R}
},
\end{equation}
where \(C_d\) is the discharge coefficient, \(\Delta p_V=p_V^{in}-p_V^{out}\) is the signed pressure difference, \(\nu_V^R\) is the specific volume at the restriction, and \(K_V(\Delta p_V)\) is an effective resistance factor. The sign of \(\Delta p_V\) determines the flow direction, while the absolute value in the denominator regularizes the pressure-drop magnitude.

To avoid numerical discontinuities near zero pressure drop, the valve model blends the laminar and turbulent limits using a pressure threshold. The limiting form of the resistance factor is:
\begin{equation}
K_V(\Delta p_V) \approx
\begin{cases}
K_{\mathrm{lam}}\Delta p_V^{\mathrm{thr}},
&
0\le |\Delta p_V| \ll \Delta p_V^{\mathrm{thr}},
\\[3pt]
K_{\mathrm{tur}}(\alpha_V,\nu_V)|\Delta p_V|,
&
|\Delta p_V|\gtrsim \Delta p_V^{\mathrm{thr}},
\end{cases}
\end{equation}
with
\begin{equation}
\Delta p_V^{\mathrm{thr}}
=
\frac{p_V^{\mathrm{in}}+p_V^{\mathrm{out}}}{2}
\left(1-B_{\mathrm{lam}}\right).
\end{equation}
Here, \(B_{\mathrm{lam}}\) is the laminar--turbulent blending parameter,
\(K_{\mathrm{lam}}\) is the laminar resistance coefficient, and
\(K_{\mathrm{tur}}\) is the turbulent resistance coefficient. The turbulent
resistance depends on the local void fraction, \(\alpha_V\), and the mixture
specific volume, \(\nu_V\), at the valve restriction. These variables account
for the two-phase nature of the flow through the restriction, including the
effect of vapor volume fraction and mixture density on the pressure-drop
relation. With the value of \(B_{\mathrm{lam}}\) listed in
Table~\ref{tab:valve_params}, the threshold becomes
\(\Delta p_V^{\mathrm{thr}}=0.001p_{V,\mathrm{avg}}\), where
\(p_{V,\mathrm{avg}}=(p_V^{\mathrm{in}}+p_V^{\mathrm{out}})/2\). Under the
operating conditions considered here, the pressure drop across the valve is
much larger than this threshold, so the valve operates predominantly in the
turbulent regime.

\subsection{Steam turbine model}\label{Sec:turbine-model}

The steam turbine was represented using the Turbine (2P) block in Simscape Fluids \cite{SimscapeFluidsR2023a}. The component acts as a two-phase expander that connects the secondary fluid network to a rotational mechanical shaft. The nominal stage conditions, efficiencies, and flow areas used for the high-pressure (HP) and low-pressure (LP) turbine stages are listed in Table~\ref{tab:turbine_params}.

\begin{table}[thbp!]
\caption{Steam-turbine stage parameters.
The HP and LP turbine stages were implemented using two-phase turbine blocks scaled from nominal operating conditions.}
\label{tab:turbine_params}
\centering
\small
\renewcommand{\arraystretch}{1.15}
\begin{tabular}{llll}
\toprule
\textbf{Parameter} & \textbf{Description} & \textbf{HP stage} & \textbf{LP stage} \\
\midrule
\(p_{T,A}^{nom}\) & Nominal inlet pressure & \(3.448~\mathrm{MPa}\) & \(1.0~\mathrm{MPa}\) \\
\(p_{T,B}^{nom}\) & Nominal outlet pressure & \(0.67~\mathrm{MPa}\) & \(0.0085~\mathrm{MPa}\) \\
\(\dot{m}_{T}^{nom}\) & Nominal mass flow rate & \(67.07~\mathrm{kg\,s^{-1}}\) & \(67.07~\mathrm{kg\,s^{-1}}\) \\
\(\nu_{T,A}^{nom}\) & Nominal inlet specific volume & \(0.07205~\mathrm{m^3\,kg^{-1}}\) & \(0.4~\mathrm{m^3\,kg^{-1}}\) \\
\(\eta_{isen}\) & Isentropic efficiency & \(0.8828\) & \(0.8828\) \\
\(\eta_{mech}\) & Mechanical efficiency & \(0.8828\) & \(0.8828\) \\
\(A_{T,A}\) & Inlet fluid flow area & \(0.08~\mathrm{m^2}\) & \(0.20~\mathrm{m^2}\) \\
\(A_{T,B}\) & Outlet fluid flow area & \(0.20~\mathrm{m^2}\) & \(6.0~\mathrm{m^2}\) \\
\bottomrule
\end{tabular}
\end{table}

For each turbine stage, conservation of mass requires:
\begin{equation}
\dot{m}_{T}^{in}+\dot{m}_{T}^{out}=0,
\end{equation}
where \(\dot{m}_{T}^{in}\) and \(\dot{m}_{T}^{out}\) are the inlet and outlet mass flow rates. The port velocities are related to the local mass flow rates by:
\begin{equation}
\dot{m}_{T}^{in}
=
\frac{A_T^{in}}{\nu_T^{in}}w_T^{in},
\qquad
\dot{m}_{T}^{out}
=
\frac{A_T^{out}}{\nu_T^{out}}w_T^{out},
\end{equation}
where \(A_T^{in}\) and \(A_T^{out}\) are the effective inlet and outlet flow areas, respectively, \(\nu_T^{in}\) and \(\nu_T^{out}\) are the corresponding specific volumes, and \(w_T^{in}\) and \(w_T^{out}\) are the local flow speeds. The use of \(\nu\) for specific volume avoids ambiguity with velocity.

With the sign convention used here, the fluid power extracted by the turbine is:

\begin{equation}
P_{\mathrm{fluid}} = \dot{m}_{T}^{in}\Delta h_T^{\mathrm{tot}},
\end{equation}
where \(\Delta h_T^{\mathrm{tot}}\) is the specific total enthalpy drop across the turbine stage. The corresponding energy balance can be written as:
\begin{equation}
\Phi_T^{in}+\Phi_T^{out}
=
P_{\mathrm{fluid}},
\end{equation}
where \(\Phi_T^{in}\) and \(\Phi_T^{out}\) are the energy flow rates through the turbine ports.

The thermodynamic closure is obtained by comparing the real expansion to an ideal isentropic expansion between the same inlet state and outlet pressure. The ideal process satisfies:
\begin{equation}
\Delta s_T^{isen}=0,
\end{equation}
and the actual enthalpy drop is computed as:
\begin{equation}
\Delta h_T^{\mathrm{tot}}
=
\eta_{isen}\Delta h_T^{\mathrm{tot, isen}},
\end{equation}
where \(\eta_{isen}\) is the isentropic efficiency and \(\Delta h_T^{isen}\) is the ideal specific enthalpy drop. Water and steam properties are evaluated from the two-phase property formulation used by the Simscape fluid network.

The turbine pressure-flow relation is governed by Stodola's ellipse:
\begin{equation}
\left(\dot{m}_{T}^{in}\right)^2
\frac{\nu_T^{in}}{p_T^{in}}
=
k_T^2
\left[
1-
\left(
\frac{p_T^{out}}{p_T^{in}}
\right)^{n_T}
\right],
\end{equation}
where \(p_T^{{in}}\) and \(p_T^{out}\) are the inlet and outlet pressures, respectively, \(k_T\) is the Stodola constant, and \(n_T\) is the polytropic exponent. This relation allows the steam flow rate to vary with the instantaneous pressure ratio rather than being imposed as a linear function of valve position or load demand.

The useful mechanical power delivered to the shaft is obtained from the extracted fluid power after mechanical losses:
\begin{equation}
P_{\mathrm{mech}}
=
M_T\omega_T^{pos}
=
\eta_{mech}P_{\mathrm{fluid}}
=
\eta_{mech}\dot{m}_{T}^{in}\Delta h_T^{\mathrm{tot}},
\end{equation}
where \(M_T\) is the shaft torque, \(\omega_T^{pos}\) is a smoothed non-negative shaft speed, and \(\eta_{mech}\) is the mechanical efficiency. The smoothed speed is defined as:
\begin{equation}
\omega_T^{pos}
=
\frac{1}{2}
\left(
\omega_T+
\sqrt{\omega_T^2+\left(\omega_T^{thr}\right)^2}
\right),
\end{equation}
where \(\omega_T^{thr}\) is a small threshold used to regularize the model near zero speed.

The same formulation was applied to the high-pressure and low-pressure turbine stages. The total turbine mechanical power is therefore:
\begin{equation}
P_{\mathrm{mech}}^{tot}
=
P_{\mathrm{mech}}^{HP}
+
P_{\mathrm{mech}}^{LP}.
\end{equation}
The turbine-stage parameters were selected to reproduce the rated operating point of the integrated SMR--Rankine model. The Stodola constants are computed from the nominal pressure, mass-flow, and inlet-specific-volume values listed in Table~\ref{tab:turbine_params}. Because \(\Delta h_T\) is recalculated from the instantaneous thermodynamic states, the turbine model captures the load-dependent specific work that is absent from the linear transfer-function baseline models.

\subsection{Condenser model}\label{Sec:condenser-model}

The condenser was represented as a rigid, constant-volume two-phase control volume in which turbine exhaust steam rejects heat and condenses into liquid water. It retains finite mass and energy storage, allowing condenser pressure, temperature, and liquid inventory to evolve dynamically during load-following transients. The condenser mixture was assumed to remain at thermodynamic saturation. The condenser volume, port areas, liquid-fraction limits, and initial liquid-fraction target are listed in Table~\ref{tab:condenser_params}.

\begin{table}[thbp!]
\caption{Condenser saturated-fluid chamber parameters.}
\label{tab:condenser_params}
\centering
\small
\renewcommand{\arraystretch}{1.15}
\begin{tabular}{lll}
\toprule
\textbf{Parameter} & \textbf{Description} & \textbf{Value} \\
\midrule
\(V_C\) & Total condenser fluid volume & \(20~\mathrm{m^3}\) \\
\(A_{C,A}\) & Cross-sectional area at vapor port A & \(6~\mathrm{m^2}\) \\
\(A_{C,B}\) & Cross-sectional area at liquid port B & \(0.03~\mathrm{m^2}\) \\
\(\psi_{C}^{\min}\) & Minimum liquid volume fraction & \(0.01\) \\
\(\psi_{C}^{\max}\) & Maximum liquid volume fraction & \(0.99\) \\
\(\psi_{C,0}\) & Initial liquid volume fraction target & \(0.5\) \\
\bottomrule
\end{tabular}
\end{table}

The total condenser mass, \(m_C\), evolves according to the integral mass balance.
\begin{equation}
\frac{\mathrm{d}m_C}{\mathrm{d}t}
=
\dot{m}_{C}^{in}
-
\dot{m}_{C}^{out},
\end{equation}
where \(\dot{m}_{C}^{in}\) and \(\dot{m}_{C}^{out}\) are the inlet and outlet mass flow rates, respectively. The corresponding energy balance is:
\begin{equation}
\frac{\mathrm{d}}{\mathrm{d}t}
\left(m_C u_C\right)
=
\dot{m}_{C}^{in}h_{C}^{\mathrm{tot,in}}
-
\dot{m}_{C}^{out}h_{C}^{\mathrm{tot,out}}
+
\dot{Q}_C ,
\end{equation}
where \(u_C\) is the mean specific internal energy of the condenser inventory, \(h_C^{\mathrm{tot,in}}\) and \(h_C^{\mathrm{tot,out}}\) are the inlet and outlet specific total enthalpies, respectively, and \(\dot{Q}_C\) is the heat-transfer rate to the condenser control volume. With this sign convention, heat rejection from the condenser corresponds to \(\dot{Q}_C<0\). The total specific enthalpy used in the convective terms includes flow work and kinetic energy.
\begin{equation}
h^{\mathrm{tot}}=u+p\nu+\frac{w^2}{2},
\end{equation}
where \(u\) is specific internal energy, \(p\) is pressure, \(\nu\) is specific volume, and \(w\) is the local flow speed.

The mean internal energy of the two-phase inventory was computed from the saturated liquid and vapor masses.
\begin{equation}
m_C u_C
=
m_C^{vap}u_v
+
m_C^{liq}u_l,
\end{equation}
where \(m_C^{vap}\) and \(m_C^{liq}\) are the vapor and liquid masses, respectively, and \(u_v\) and \(u_l\) are the saturated vapor and liquid internal energies evaluated at the instantaneous condenser pressure \(p_C\).

The fixed condenser volume, \(V_C\), was partitioned into vapor and liquid regions using the liquid volume fraction \(\psi_C\).
\begin{equation}
\psi_C
=
\frac{V_{liq}}{V_C},
\qquad
0\leq \psi_C \leq 1 .
\end{equation}
The corresponding vapor and liquid masses are:
\begin{equation}
m_C^{vap}
=
\frac{(1-\psi_C)V_C}{\nu_{vap}(p_C)},
\qquad
m_C^{liq}
=
\frac{\psi_C V_C}{\nu_{liq}(p_C)},
\end{equation}
and the total inventory is:
\begin{equation}
m_C=m_C^{vap}+m_C^{liq}.
\end{equation}
Here, \(\nu_{vap}(p_C)\) and \(\nu_{liq}(p_C)\) are the saturated vapor and liquid specific volumes at the condenser pressure, respectively. The liquid volume fraction \(\psi_C\) was used as the condenser-inventory indicator in the transient analysis.

\subsection{Feedwater pump model}\label{Sec:pump-model}

The feedwater pump was represented as a controlled mass-flow source in the two-phase Simscape network. This choice treats the pump as a robust actuator for plant-level control studies. The commanded feedwater flow is imposed directly, while the surrounding two-phase network determines the pressure rise, outlet state, enthalpy change, and mechanical power requirement. The pump was implemented with equal inlet and outlet port areas of \(0.03~\mathrm{m^2}\), and the commanded mass flow was supplied by the feedwater control loop.

\begin{equation}
\dot{m}_{fw}^{in}
=
\dot{m}_{fw}^{cmd},
\end{equation}
and conservation of mass gives:
\begin{equation}
\dot{m}_{fw}^{in}
+
\dot{m}_{fw}^{out}
=
0 ,
\end{equation}
where \(\dot{m}_{fw}^{cmd}\) is the command generated by the feedwater control system, and \(\dot{m}_{fw}^{in}\) and \(\dot{m}_{fw}^{out}\) are the inlet and outlet mass flow rates of the pump, respectively.

The pump exchanges energy with the fluid through mechanical work. The steady-flow energy balance is:
\begin{equation}
\dot{m}_{fw}^{in}h_{fw}^{\mathrm{tot,in}}
+
\dot{m}_{fw}^{out}h_{fw}^{\mathrm{tot,out}}
+
P_{fw}^{ext}
=
0,
\end{equation}
where \(h_{fw}^{\mathrm{tot,in}}\) and \(h_{fw}^{\mathrm{tot,out}}\) are the inlet and outlet total specific enthalpies, respectively, and \(P_{fw}^{ext}\) is the external mechanical power supplied to the pump. Using \(\dot{m}_{fw}^{out}=-\dot{m}_{fw}^{in}\), the mechanical power input can be written as:
\begin{equation}
P_{fw}^{ext}
=
\dot{m}_{fw}^{in}
\left(
h_{fw}^{\mathrm{tot,out}}
-
h_{fw}^{\mathrm{tot,in}}
\right).
\end{equation}

The outlet state was obtained by idealizing the pressure-changing process across the pump as internally reversible and adiabatic. The entropy is therefore conserved across the pump:
\begin{equation}
s_{\mathrm{fw}}^{\mathrm{out}}
=
s_{\mathrm{fw}}^{\mathrm{in}} .
\end{equation}
The outlet enthalpy is evaluated from the outlet pressure imposed by the fluid network and the inlet entropy:
\begin{equation}
h_{\mathrm{fw}}^{\mathrm{out}}
=
h\left(
p_{\mathrm{fw}}^{\mathrm{out}},
s_{\mathrm{fw}}^{\mathrm{in}}
\right),
\end{equation}
where \(h(p,s)\) denotes the water/steam property relation used to evaluate specific enthalpy from pressure and entropy. Thus, while the pump imposes the commanded feedwater mass flow, the outlet thermodynamic state and required mechanical power are determined consistently with the surrounding two-phase network.

Together, these component models define the physics-based Rankine-cycle network used in the integrated SMR--steam-cycle simulation. The next section describes how this acausal two-phase network is coupled to the equation-based SMR and moving-boundary SG model, and how the valve, feedwater pump, and control rods are coordinated during load-following transients.

\section{Integrated SMR--Rankine coupling and control}\label{Sec:integrated-control}

\subsection{Simulink--Simscape thermodynamic coupling interface}\label{Sec:coupling-interface}

The equation-based SMR and SG model is implemented in MATLAB/Simulink, whereas the secondary Rankine-cycle balance of plant is implemented using two-phase Simscape Fluids components. The SG model computes the secondary-side pressure and boundary temperatures but does not expose physical two-phase fluid ports. In contrast, the steam throttle valve, turbine, condenser, and feedwater pump are connected through acausal two-phase ports. To couple these domains without imposing a purely causal steam-flow relation, controlled two-phase reservoirs are used as thermodynamic interfaces between the Simulink SG states and the Simscape secondary network.

The hot-side reservoir represents the SG outlet boundary. It imposes the SG secondary pressure and superheated outlet temperature on the valve inlet:
\begin{equation}
p_R^{\mathrm{sh}}=p_s,
\qquad
T_R^{\mathrm{sh}}=T_{\mathrm{SG}}^{\mathrm{out}},
\qquad
\dot{m}_R^{\mathrm{out}}=\dot{m}_V^{\mathrm{in}}=\dot{m}_s^{\mathrm{out}} .
\end{equation}
The cold-side reservoir represents the SG inlet boundary. It accepts the feedwater pump discharge and imposes the SG inlet state on the secondary network:
\begin{equation}
p_R^{\mathrm{sc}}=p_s,
\qquad
T_R^{\mathrm{sc}}=T_{\mathrm{SG}}^{\mathrm{in}},
\qquad
\dot{m}_R^{\mathrm{in}}=\dot{m}_{\mathrm{fw}}^{\mathrm{out}}=\dot{m}_s^{\mathrm{in}} .
\end{equation}
Here, \(p_s\) is the uniform secondary-side SG pressure, 
\(T_{\mathrm{SG}}^{\mathrm{out}}\) is the SG outlet temperature, 
\(T_{\mathrm{SG}}^{\mathrm{in}}\) is the feedwater inlet temperature, 
\(\dot{m}_s^{\mathrm{out}}\) is the steam flow leaving the SG, and 
\(\dot{m}_s^{\mathrm{in}}\) is the feedwater flow entering the SG. 
Superscripts \(\mathrm{sh}\) and \(\mathrm{sc}\) denote the superheated and 
subcooled interface reservoirs, respectively.

Thermodynamic-property consistency across the Simulink--Simscape interface was maintained by configuring the equation-based SMR--SG model and the Simscape two-phase Rankine-cycle network with consistent water/steam property data. This prevents artificial discontinuities in density, enthalpy, or saturation state across the controlled-reservoir interface and ensures that the pressure, temperature, and flow variables exchanged between the equation-based SG model and the physics-based Rankine-cycle network remain thermodynamically compatible.

For each imposed pressure--temperature pair, the reservoir evaluates the corresponding water/steam properties, including specific internal energy \(u_R\), specific volume \(\nu_R\), and specific enthalpy \(h_R\).

The interface reservoirs do not introduce additional dynamic storage. Instead, they source or sink the mass flow demanded by the connected two-phase components. The port mass flow and energy flux are evaluated as
\begin{equation}
\dot{m}_R
=
\frac{A_R}{\nu_R}w_R,
\end{equation}
and
\begin{equation}
\Phi_R
=
\dot{m}_R
\left(
h_R+\frac{w_R^2}{2}
\right),
\end{equation}
where \(A_R\) is the port flow area and \(w_R\) is the local flow speed. This interface allows the Rankine-cycle network to impose pressure--flow constraints and back-pressure effects on the SG while preserving the equation-based moving-boundary formulation of the SMR module.

\begin{figure}[htbp!]
  \centering
  \includegraphics[width=\linewidth]{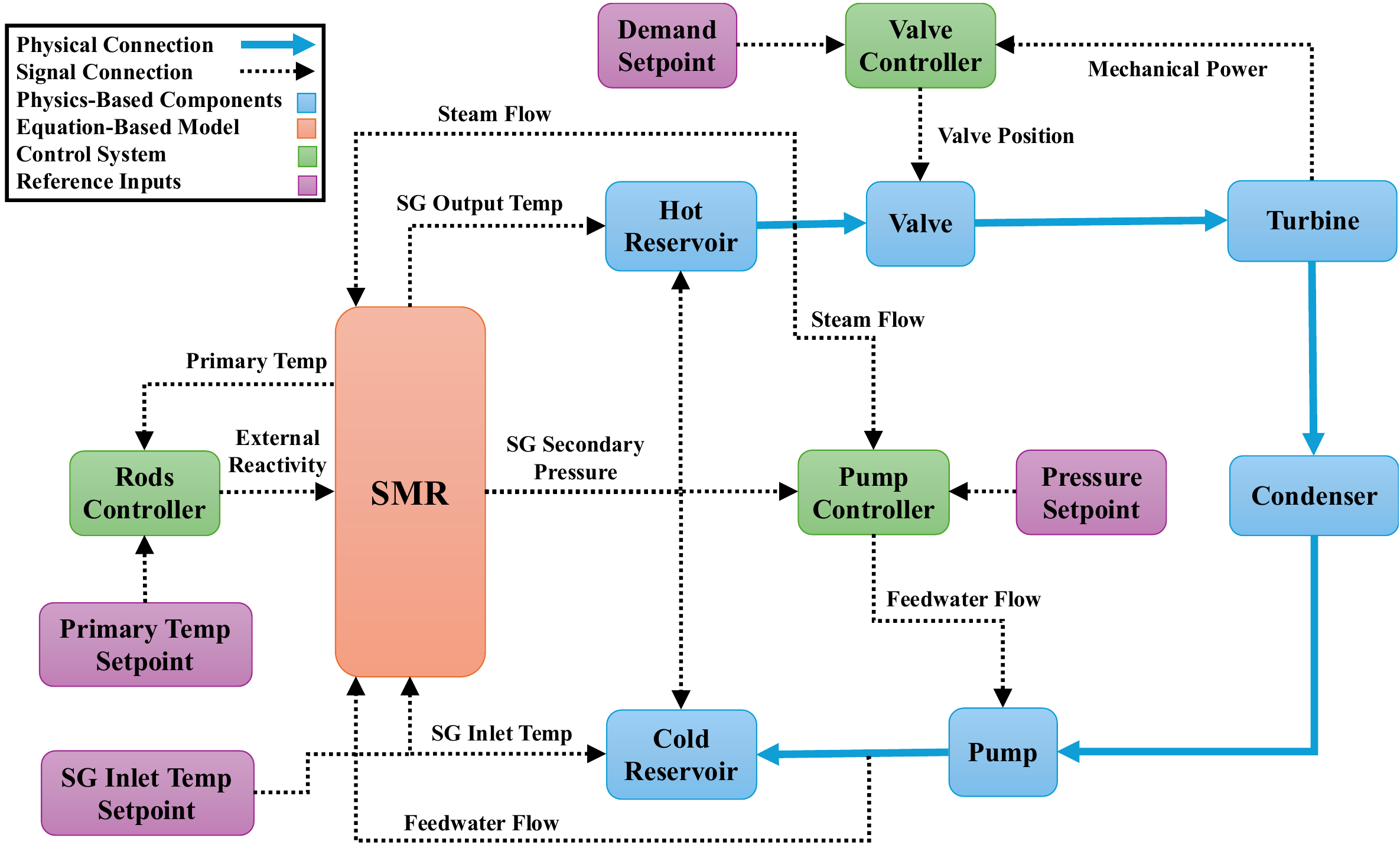}
  \caption{Integrated SMR--Rankine modeling and control architecture. The equation-based SMR and moving-boundary SG model is coupled to a physics-based secondary Rankine-cycle network through controlled two-phase reservoirs. Dotted arrows denote causal signal connections for measurements, references, and actuator commands, while solid connections denote acausal two-phase physical connections that enforce mass and energy conservation in the Rankine-cycle network.}
  \label{fig:Block_diagram_Physics-based_model}
\end{figure}

Figure~\ref{fig:Block_diagram_Physics-based_model} summarizes the integrated modeling and control architecture. The framework is divided into four domains: the equation-based SMR and SG model, the physics-based Rankine-cycle network, the load-following controllers, and the external reference inputs. Signal connections represent causal information transfer, including measured variables sent to the controllers and actuator commands sent to the plant. Physical connections represent acausal two-phase fluid ports through which mass and energy are solved bidirectionally in the Rankine-cycle network. The hot- and cold-side controlled reservoirs form the interface between these domains by mapping the SG pressure and boundary temperatures from the equation-based model into thermodynamic states for the Simscape two-phase network. This structure allows back-pressure and pressure--flow constraints from the Rankine cycle to affect the moving-boundary SG without replacing the SG model by a purely causal steam-flow relation.

\subsection{Load-following control architecture}\label{Sec:control-architecture}

The load-following control system is implemented as a decentralized reactor-following-turbine architecture with three feedback loops. The steam throttle valve regulates turbine mechanical power, the feedwater pump regulates SG pressure, and the control rods regulate the average primary coolant temperature. These loops act on different time scales and are used to evaluate the relative importance of mechanical, hydraulic, and neutronic--thermal control during load-following transients. In this architecture, the turbine-side controller responds directly to mechanical-power demand, while the feedwater and control-rod loops restore secondary pressure and primary thermal conditions over slower plant time scales.

\subsubsection{Primary-temperature control via control rods}\label{Sec:rod-control}

The control-rod loop regulates the average primary coolant temperature, \(\bar{T}_c\), by adjusting the externally imposed reactivity \(\rho_{\mathrm{ext}}\). The temperature error is:
\begin{equation}
e_r(t)=\bar{T}_{c,\mathrm{ref}}-\bar{T}_c(t),
\end{equation}
and the reactivity command is generated by a PI controller:
\begin{equation}
\rho_{\mathrm{ext}}(t)
=
K_{P,\mathrm{rod}}
\left[
e_r(t)
+
K_{I,\mathrm{rod}}
\int_0^t e_r(\tau)\,\mathrm{d}\tau
\right].
\end{equation}
The controller gains are \(K_{P,\mathrm{rod}}=3.5\times10^{-4}\) and \(K_{I,\mathrm{rod}}=0.017~\mathrm{s}^{-1}\). A positive temperature error corresponds to a primary coolant temperature below its reference and therefore to rod withdrawal, whereas a negative temperature error corresponds to rod insertion.


\subsubsection{SG pressure and feedwater-flow control}\label{Sec:feedwater-control}

Motivated by prior SG control studies that regulate steam pressure while reducing
steam--feedwater flow mismatch~\cite{Sabir2021LoadFrequency}, the feedwater-pump loop
uses a two-element strategy that combines steam-flow feed-forward with SG pressure
feedback. The pressure error is
\begin{equation}
e_p(t)=p_{s,\mathrm{ref}}-p_s(t),
\end{equation}
where \(p_{s,\mathrm{ref}}\) is the SG pressure setpoint. The commanded feedwater flow is
\begin{equation}
\dot{m}_{\mathrm{fw}}^{\mathrm{cmd}}(t)
=
\dot{m}_s^{\mathrm{out}}(t)
+
K_{P,\mathrm{fw}}
\left[
e_p(t)
+
K_{I,\mathrm{fw}}
\int_0^t e_p(\tau)\,\mathrm{d}\tau
\right].
\end{equation}
The first term tracks the measured steam outflow, reducing sustained inlet--outlet flow
mismatch. The feedback term corrects pressure deviations by temporarily allowing
\(\dot{m}_{\mathrm{fw}}\neq\dot{m}_s^{\mathrm{out}}\), which redistributes inventory within the SG and
secondary loop during transients. The controller gains are
\(K_{P,\mathrm{fw}}=4.354\times10^{-5}\) and \(K_{I,\mathrm{fw}}=10^{-4}~\mathrm{s}^{-1}\).


\subsubsection{Turbine-power control via steam throttle valve}\label{Sec:valve-control}

The steam throttle valve regulates turbine mechanical power by adjusting the valve restriction area \(A_V^R\). The mechanical-power error is
\begin{equation}
e_v(t)=P_{\mathrm{demand}}(t)-P_{\mathrm{mech}}(t),
\end{equation}
and the PI controller output is
\begin{equation}
u_{\mathrm{PI}}(t)
=
K_{P,V}
\left[
e_v(t)
+
K_{I,V}
\int_0^t e_v(\tau)\,\mathrm{d}\tau
\right].
\end{equation}
To represent actuator dynamics, the area correction \(\Delta A_V\) follows a first-order lag:
\begin{equation}
\tau_V
\frac{\mathrm{d}\Delta A_V}{\mathrm{d}t}
+
\Delta A_V
=
u_{\mathrm{PI}}(t),
\end{equation}
and the commanded valve area is
\begin{equation}
A_V^R(t)=A_{V,\mathrm{rated}}+\Delta A_V(t).
\end{equation}
The valve-controller gains are \(K_{P,V}=0.0015\), \(K_{I,V}=2.5~\mathrm{s}^{-1}\), and the actuator time constant is \(\tau_V=0.1~\mathrm{s}\). Opening the valve increases the restriction area and therefore increases the steam flow entering the turbine, while closing the valve reduces steam flow and mechanical power.

\section{Simulation results and discussion}\label{Sec:results}

The dynamic response of the integrated SMR--Rankine model was evaluated using time-domain simulations initialized at the rated operating point. The main disturbance considered is a near-nominal load-following maneuver in which the turbine mechanical-power demand is reduced by 5\%, from \(50~\mathrm{MW}\) to \(47.5~\mathrm{MW}\), at \(t=100~\mathrm{s}\). This disturbance is used to evaluate how different actuator combinations affect mechanical-power tracking, secondary-pressure regulation, primary-loop thermal response, and steam-generator (SG) moving-boundary behavior.

\subsection{Steady-state validation at rated power}\label{Sec:steady-validation}

Before analyzing transient behavior, the integrated model was verified against the rated operating point reported in the design documentation \cite{nuscale2020dca}. Table~\ref{tab:validation} compares key steady-state quantities, including temperatures, pressures, mass flow rates, and SG operating conditions. The model reproduces the nominal design point closely, including the rated reactor thermal power, hot- and cold-leg temperatures, primary and secondary pressures, steam outlet temperature, and mass flow rates.

\begin{table}[thbp!]
\caption{Comparison of design data and simulation results at rated power.}
\label{tab:validation}
\centering
\renewcommand{\arraystretch}{1.2}
\begin{tabular}{lcc}
\toprule
\textbf{Parameter} & \textbf{Design data} & \textbf{Simulation} \\
\midrule
Rated reactor thermal power (\(\mathrm{MW}\)) & 160 & 160 \\
Hot-leg temperature (\(^{\circ}\mathrm{C}\)) & 310.06 & 310 \\
Cold-leg temperature (\(^{\circ}\mathrm{C}\)) & 258.11 & 258.11 \\
Primary pressure (\(\mathrm{MPa}\)) & 12.76 & 12.76 \\
Steam pressure (\(\mathrm{MPa}\)) & 3.448 & 3.448 \\
Steam outlet temperature (\(^{\circ}\mathrm{C}\)) & 306.88 & 306.86 \\
Feedwater inlet temperature (\(^{\circ}\mathrm{C}\)) & 148.72 & 148.72 \\
Primary mass flow rate (\(\mathrm{kg\,s^{-1}}\)) & 587 & 587 \\
Steam mass flow rate (\(\mathrm{kg\,s^{-1}}\)) & 67.07 & 67.07 \\
Steam-generator tube length (\(\mathrm{m}\)) & 24.2 & 24.2 \\
Subcooled-region length, \(L_1\) (\(\mathrm{m}\)) & -- & 2.95 \\
Two-phase-region length, \(L_2\) (\(\mathrm{m}\)) & -- & 18.49 \\
Superheated-region length, \(L_3\) (\(\mathrm{m}\)) & -- & 2.76 \\
\bottomrule
\end{tabular}
\end{table}

\begin{figure}[thbp!]
  \centering
  \includegraphics[width=0.75\linewidth]{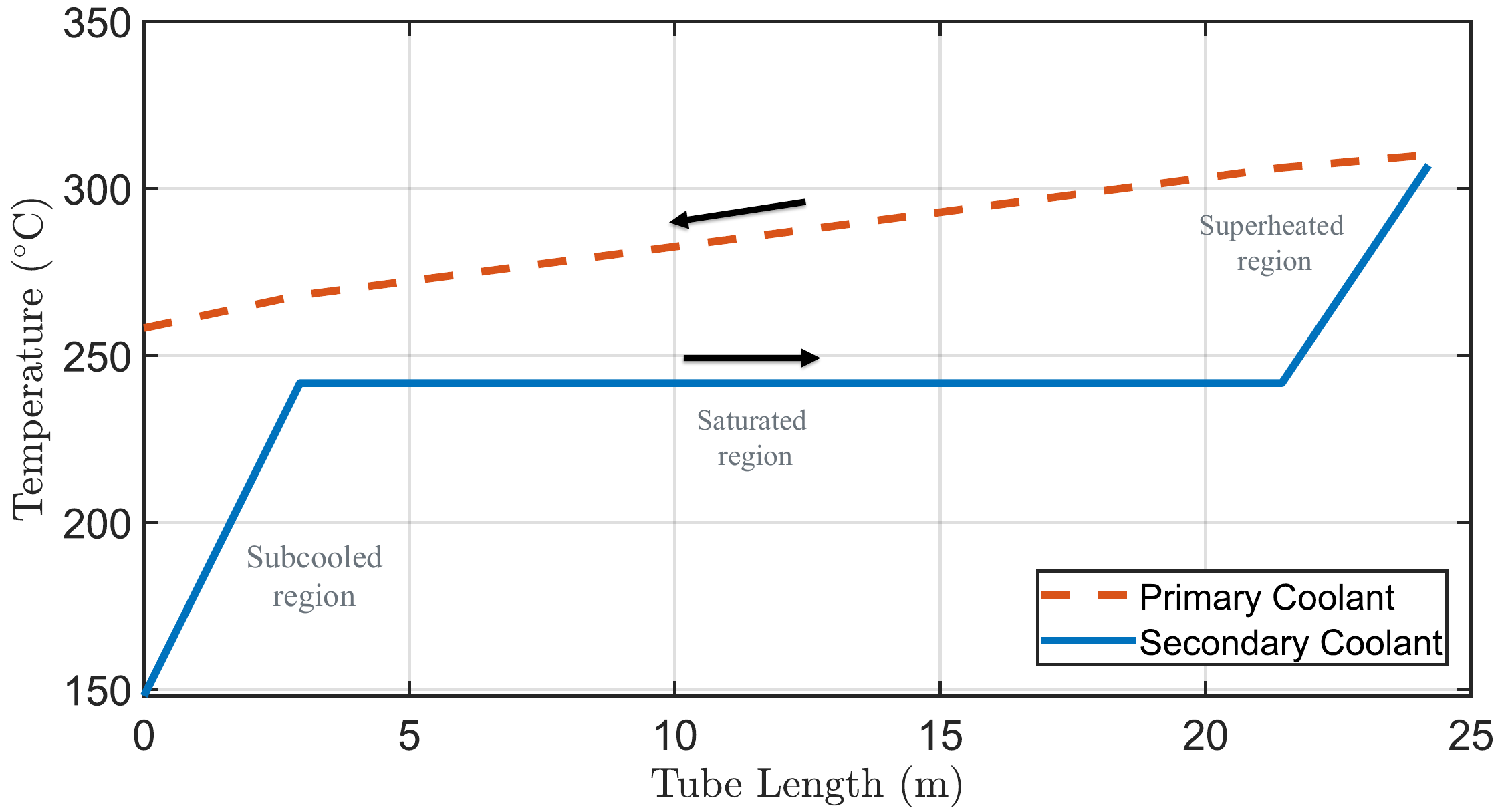}
  \caption{Steam-generator temperature profile at rated operating conditions. The moving-boundary formulation partitions the secondary side into subcooled, two-phase, and superheated regions while the primary coolant flows countercurrently on the shell side.}
  \label{fig:SG_profile}
\end{figure}

Figure~\ref{fig:SG_profile} shows the corresponding SG temperature profile along the tube length. The secondary fluid is heated from subcooled liquid to saturated two-phase flow and then to dry or superheated steam, while the primary coolant transfers heat in counterflow. Although the design documentation does not report the internal moving-boundary lengths, the simulated values of \(L_1\), \(L_2\), and \(L_3\) provide a physically consistent three-region partitioning for the rated condition and establish the initial state for the transient simulations.

\subsection{Load-following disturbance and control scenarios}\label{Sec:control-scenarios}

To isolate the roles of the valve, feedwater pump, and control rods, five control configurations were simulated under the same 5\% load-reduction disturbance:
\begin{enumerate}
    \item \textbf{Scenario 1 (No control):} An open-loop baseline in which the steam valve area is manually stepped down. The feedwater pump operates in feed-forward mode by tracking steam flow, and the control rods remain fixed.

    \item \textbf{Scenario 2 (Valve only):} The steam valve controller regulates turbine mechanical power. The feedwater pump remains in feed-forward mode, and the control rods remain fixed.

    \item \textbf{Scenario 3 (Valve + pump):} The steam valve controller regulates turbine mechanical power, and the feedwater pump uses steam-flow feed-forward together with SG pressure feedback. The control rods remain fixed.

    \item \textbf{Scenario 4 (Valve + rods):} The steam valve controller regulates turbine mechanical power, and the control rods regulate the average primary coolant temperature. The feedwater pump remains in feed-forward mode.

    \item \textbf{Scenario 5 (Valve + pump + rods):} The steam valve, feedwater pump, and control rods all operate with their corresponding feedback loops, providing the fully integrated control configuration.
\end{enumerate}

\begin{figure}[thbp!]
  \centering
  
  \subfloat{\includegraphics[width=0.48\textwidth]{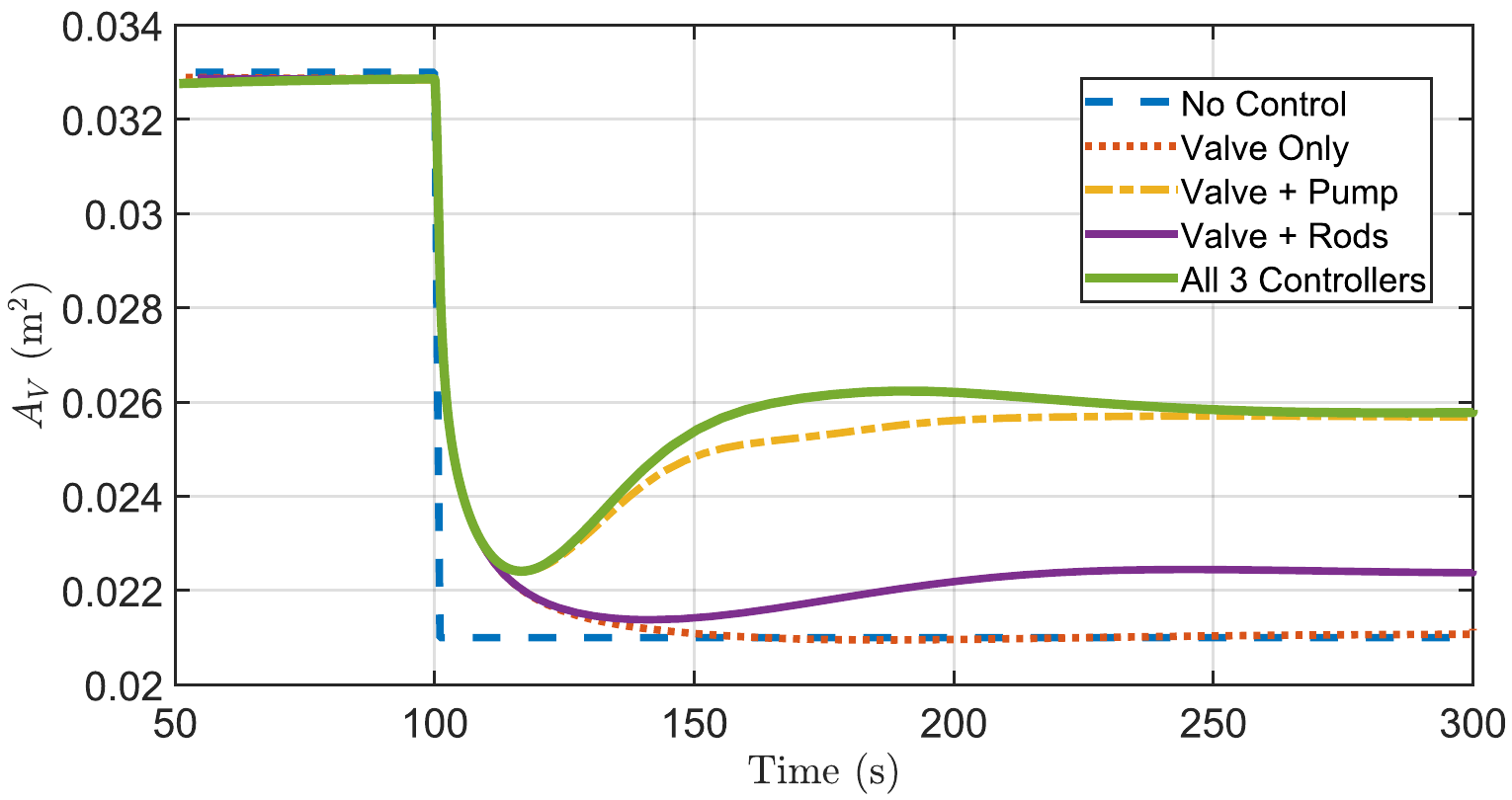}}
  \put(-215,125){\textbf{a}}
  \hspace{1em}
  \subfloat{\includegraphics[width=0.48\textwidth]{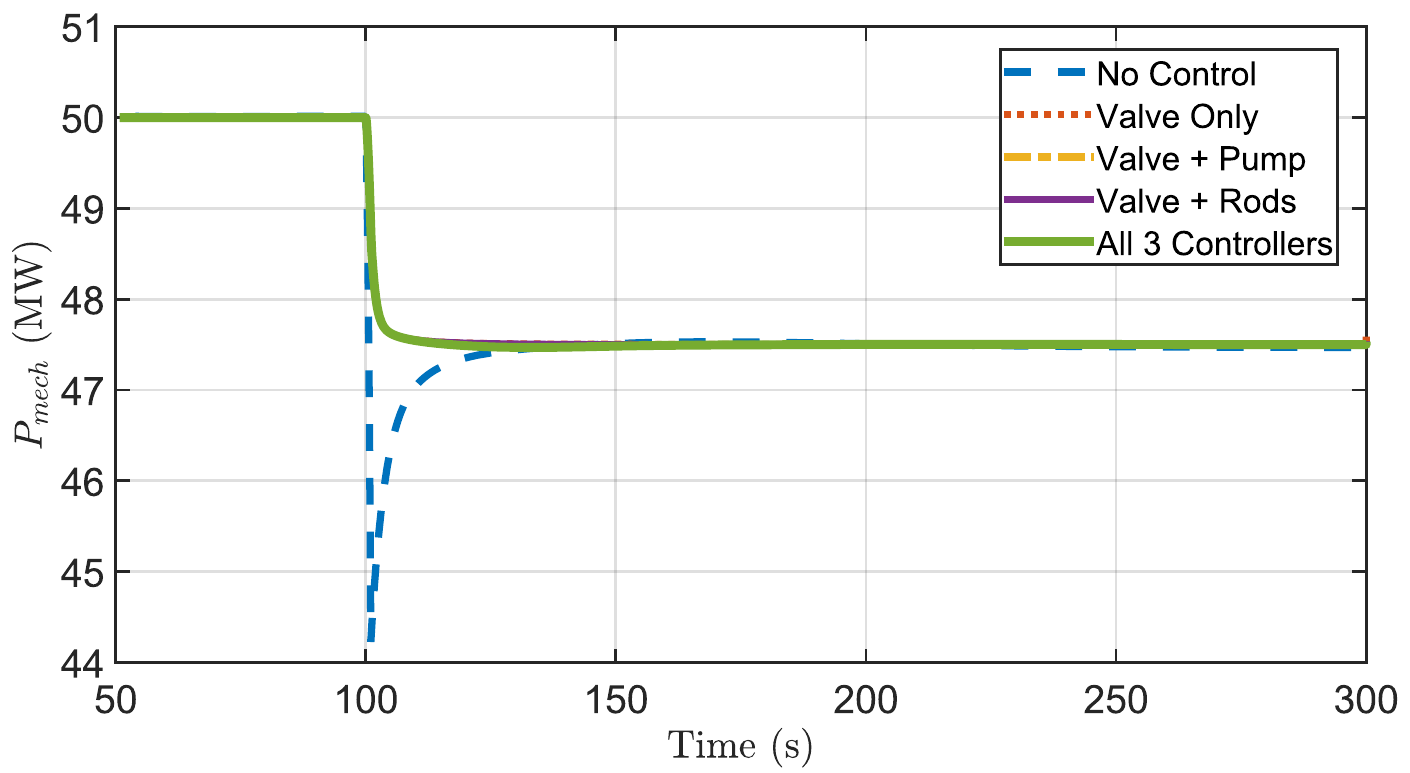}}
  \put(-215,125){\textbf{b}}
 
  \subfloat{\includegraphics[width=0.48\textwidth]{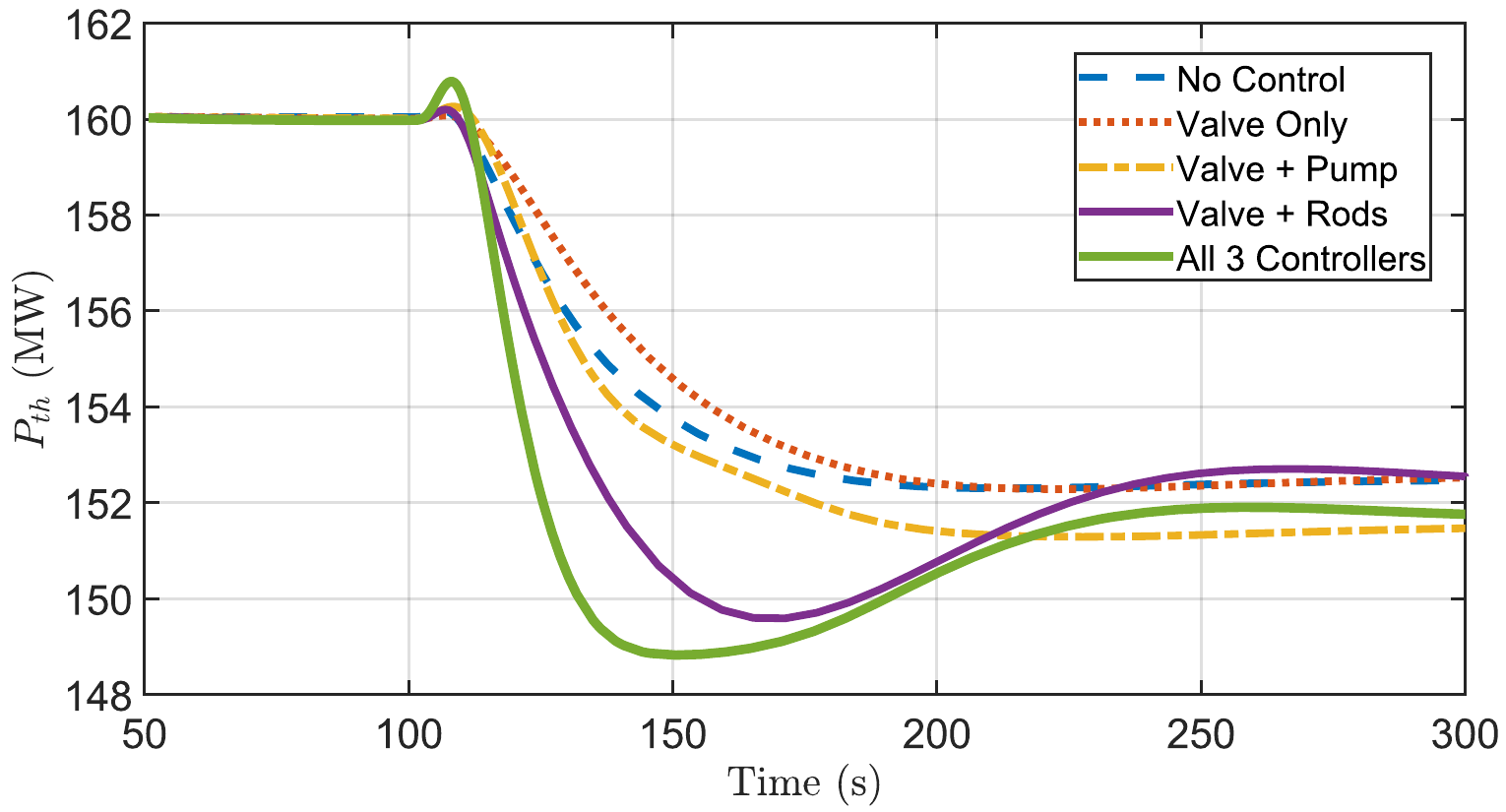}}
  \put(-215,125){\textbf{c}}
  \hspace{1em}
  \subfloat{\includegraphics[width=0.48\textwidth]{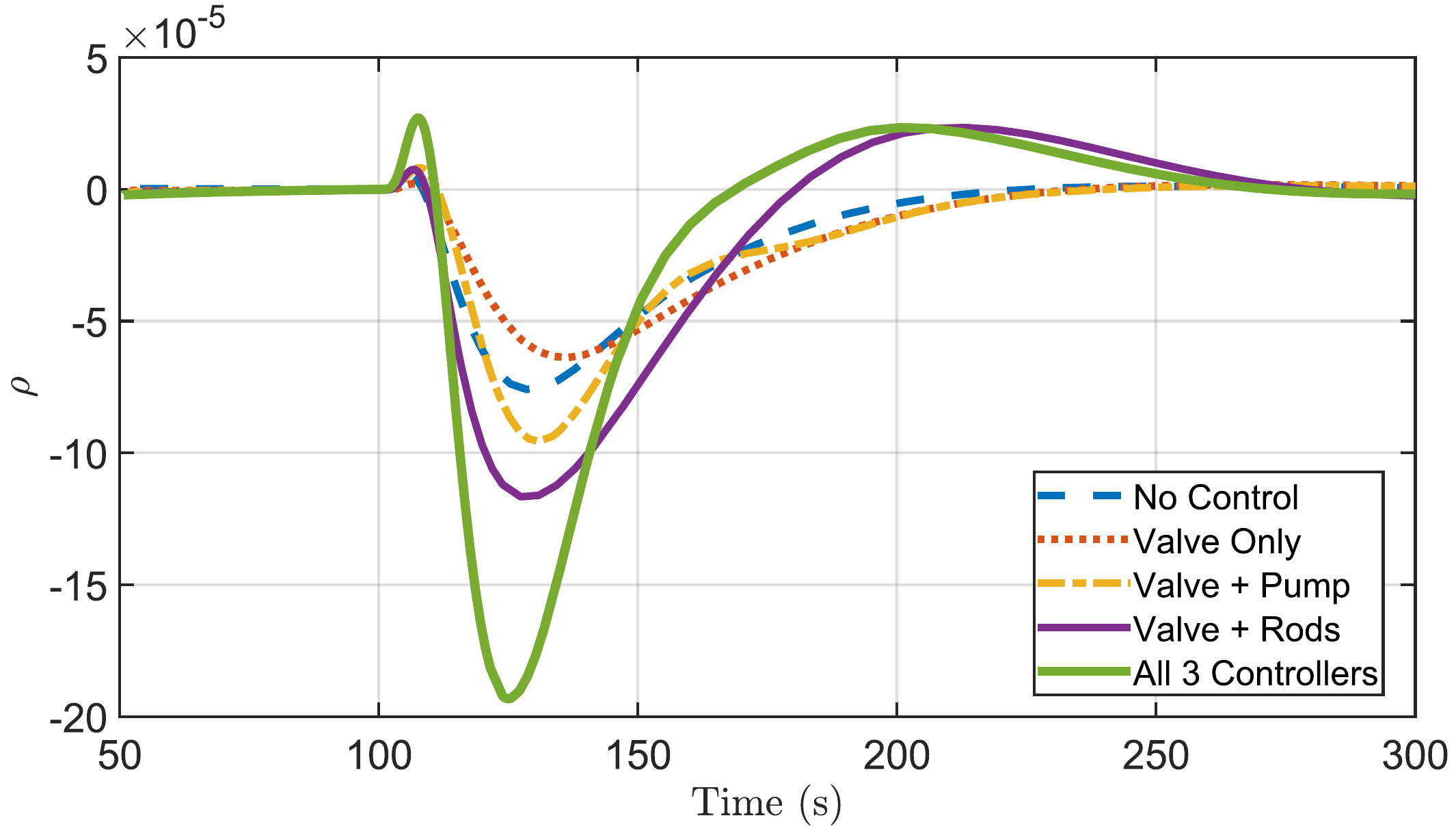}}
  \put(-215,125){\textbf{d}}

  \subfloat{\includegraphics[width=0.48\textwidth]{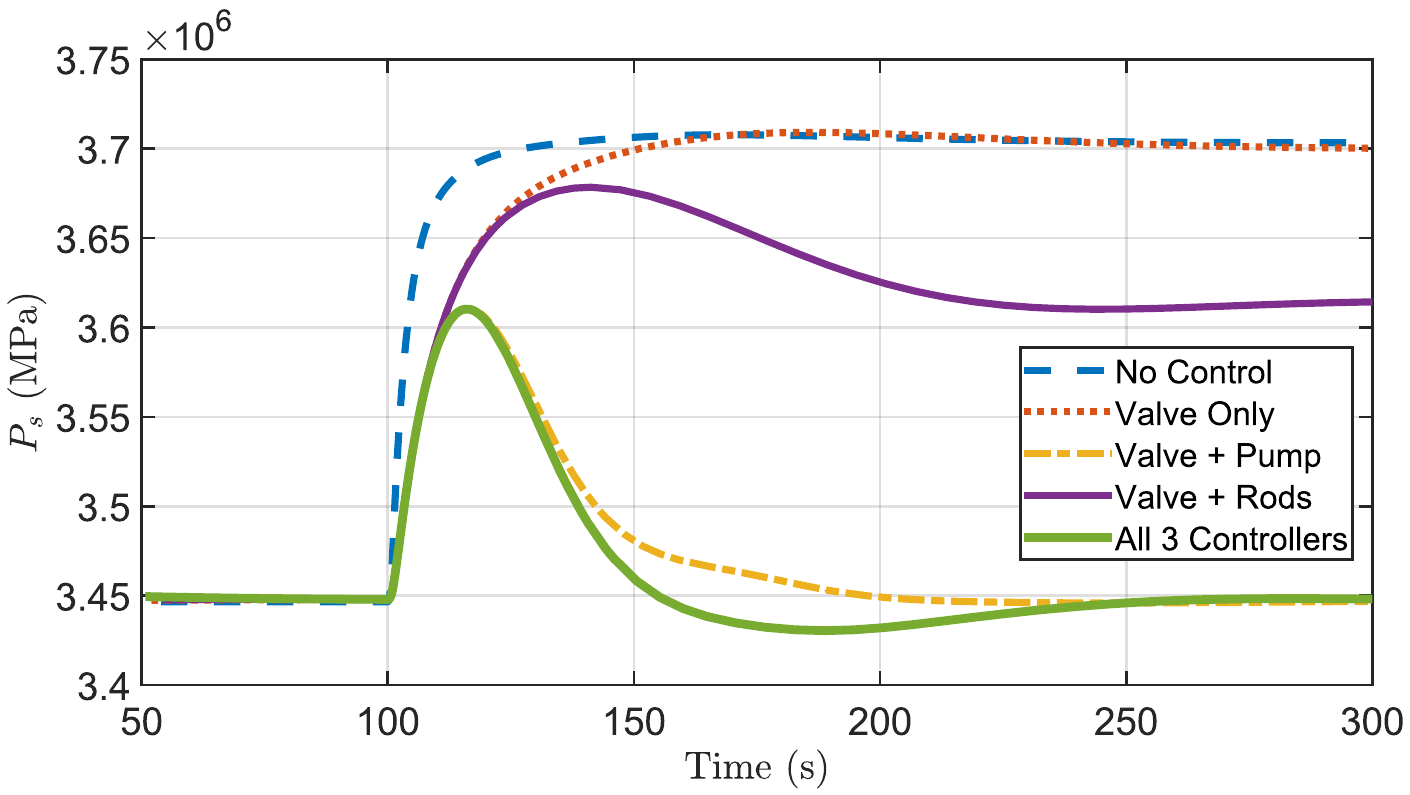}}
  \put(-215,125){\textbf{e}}
  \hspace{1em}
  \subfloat{\includegraphics[width=0.48\textwidth]{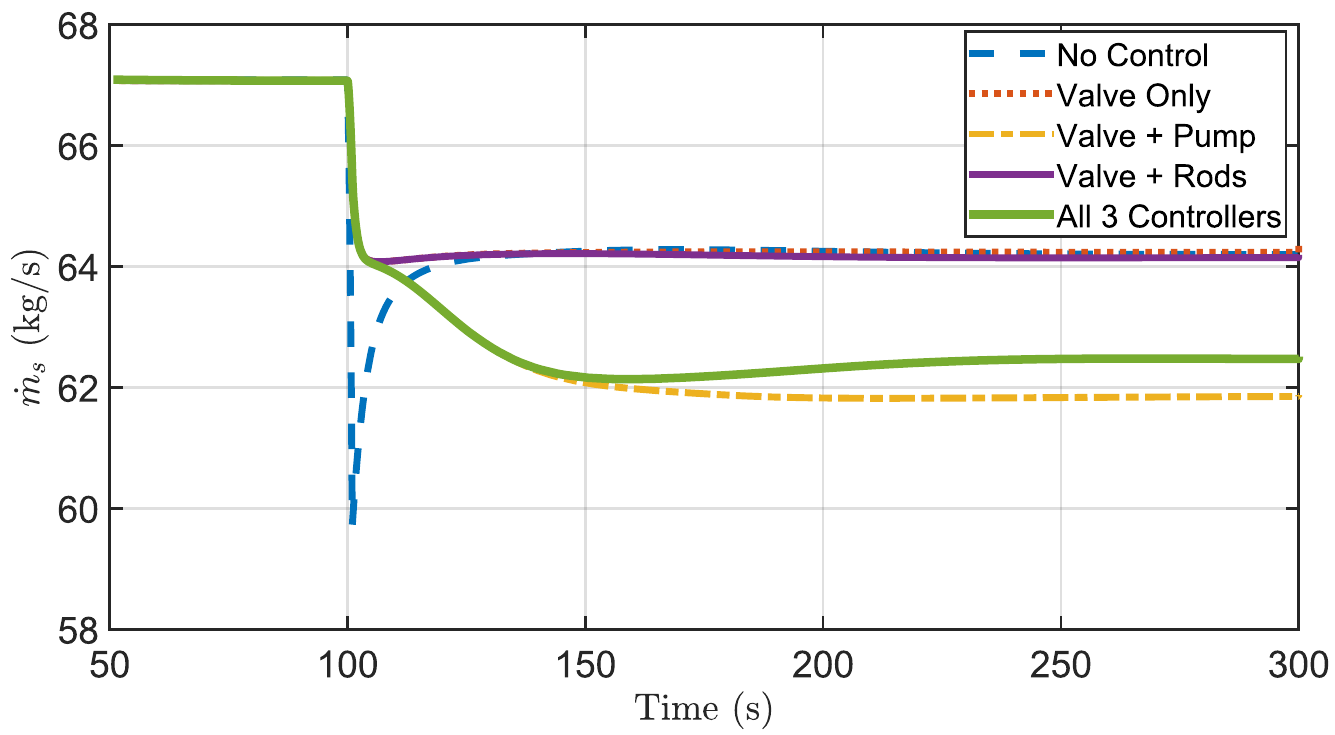}}
  \put(-215,125){\textbf{f}}

  \subfloat{\includegraphics[width=0.48\textwidth]{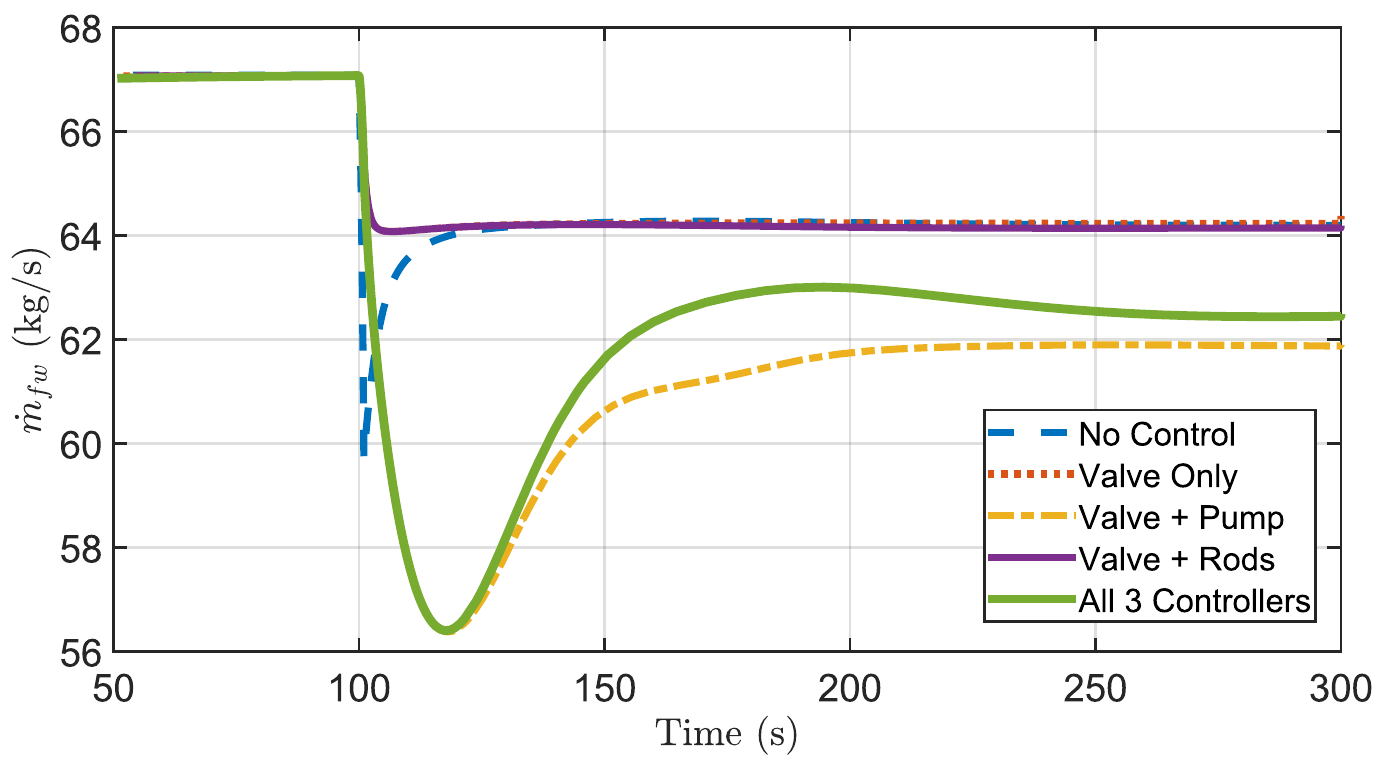}}
  \put(-215,130){\textbf{g}}
  \hspace{1em}
  \subfloat{\includegraphics[width=0.48\textwidth]{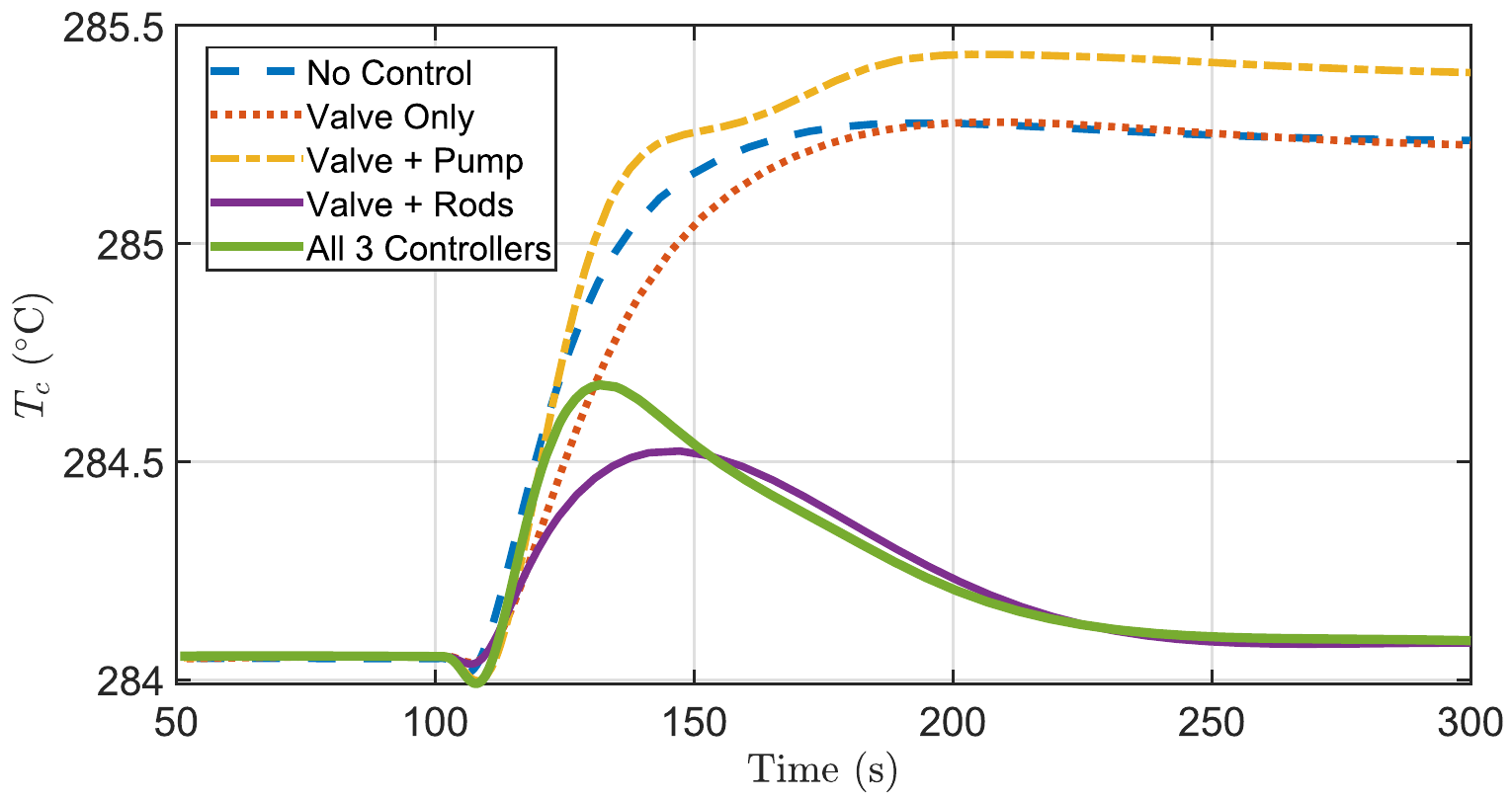}}
  \put(-215,125){\textbf{h}}
  
  \caption{Effects of decentralized control loops on SMR load-following dynamics. A 5\% step reduction in mechanical-power demand, from \(50~\mathrm{MW}\) to \(47.5~\mathrm{MW}\), is applied at \(t=100~\mathrm{s}\) under the five control configurations described in Section~\ref{Sec:control-scenarios}. Panels show: \textbf{a,} throttle-valve restriction area \(A_V^R\); \textbf{b,} turbine mechanical power \(P_{\mathrm{mech}}\); \textbf{c,} reactor thermal power \(P_{\mathrm{th}}\); \textbf{d,} core reactivity \(\rho_{\mathrm{react}}\); \textbf{e,} SG secondary pressure \(p_s\); \textbf{f,} steam mass flow rate \(\dot{m}_s\); \textbf{g,} feedwater mass flow rate \(\dot{m}_{\mathrm{fw}}\); and \textbf{h,} primary coolant average temperature \(\bar{T}_c\).}
  \label{fig:scenario_set1}
\end{figure}

\begin{figure}[thbp!]
  \centering
  
  \subfloat{\includegraphics[width=0.48\textwidth]{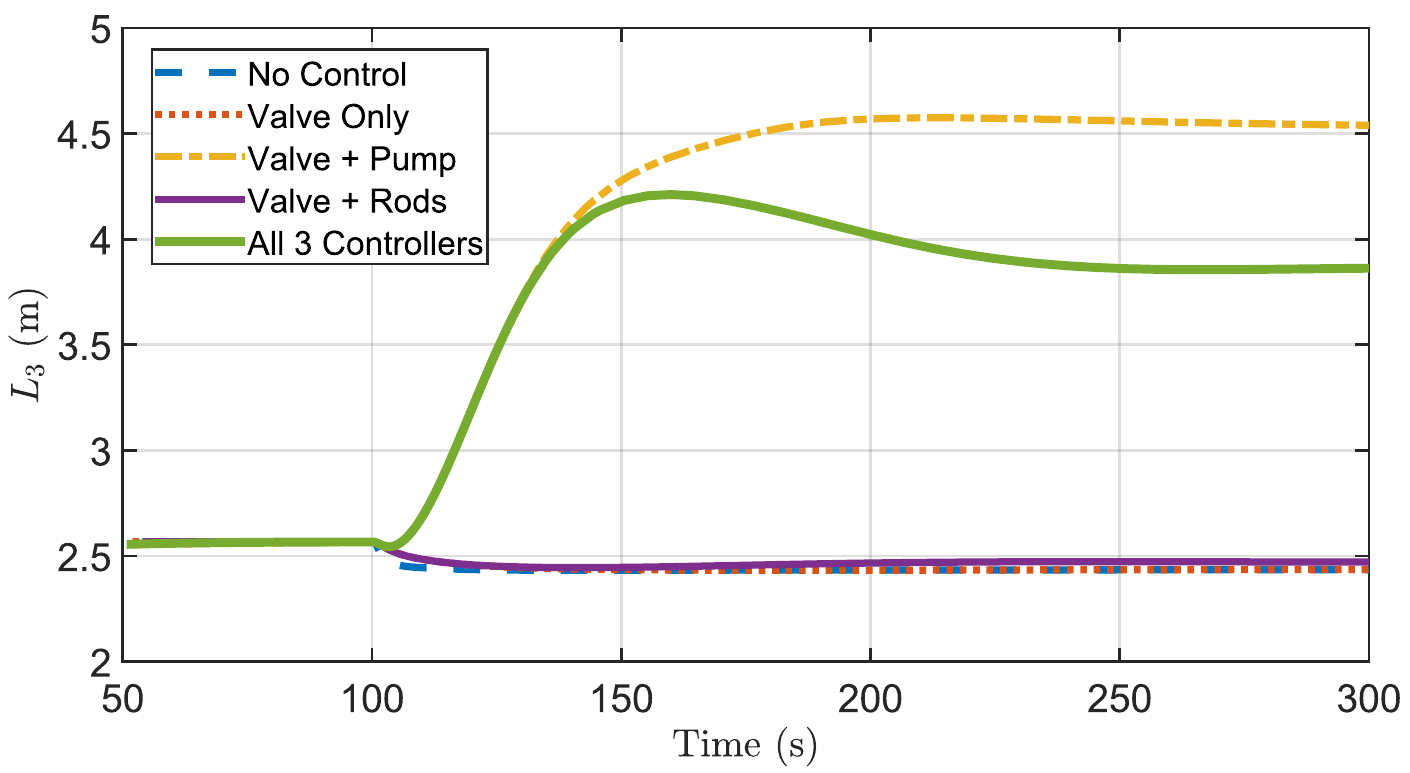}}
  \put(-215,125){\textbf{a}}
  \hspace{1em}
  \subfloat{\includegraphics[width=0.48\textwidth]{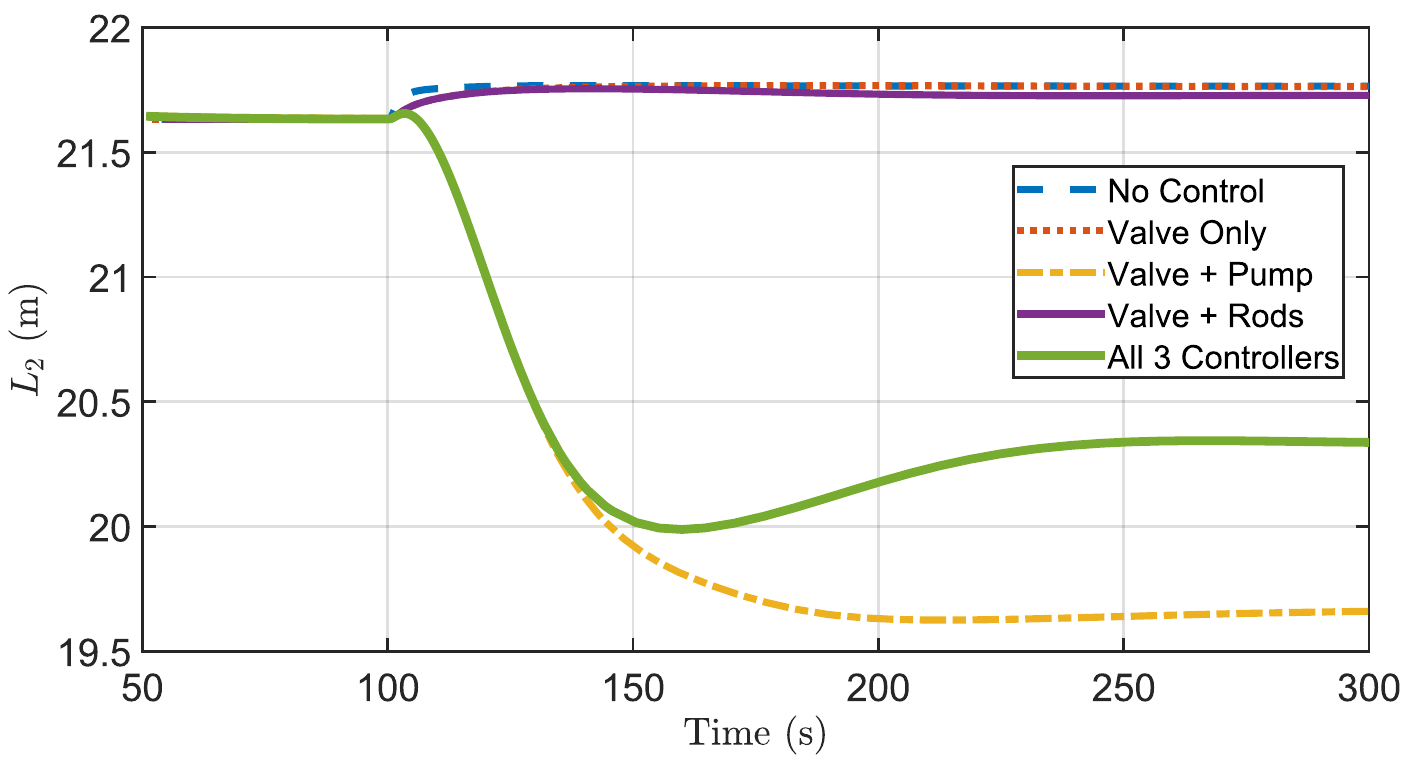}}
  \put(-215,125){\textbf{b}}
  
  \subfloat{\includegraphics[width=0.48\textwidth]{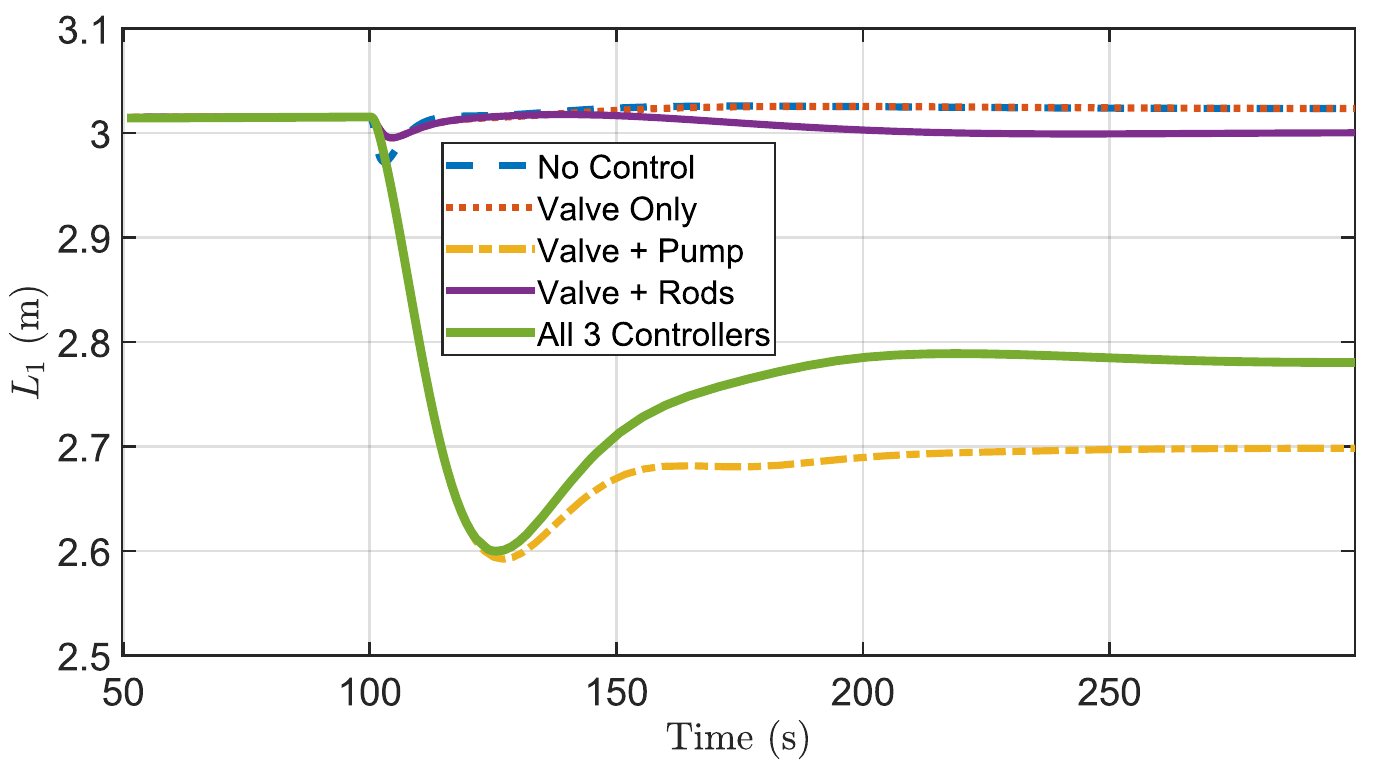}}
  \put(-215,127){\textbf{c}}  
  \hspace{1em}
  \subfloat{\includegraphics[width=0.48\textwidth]{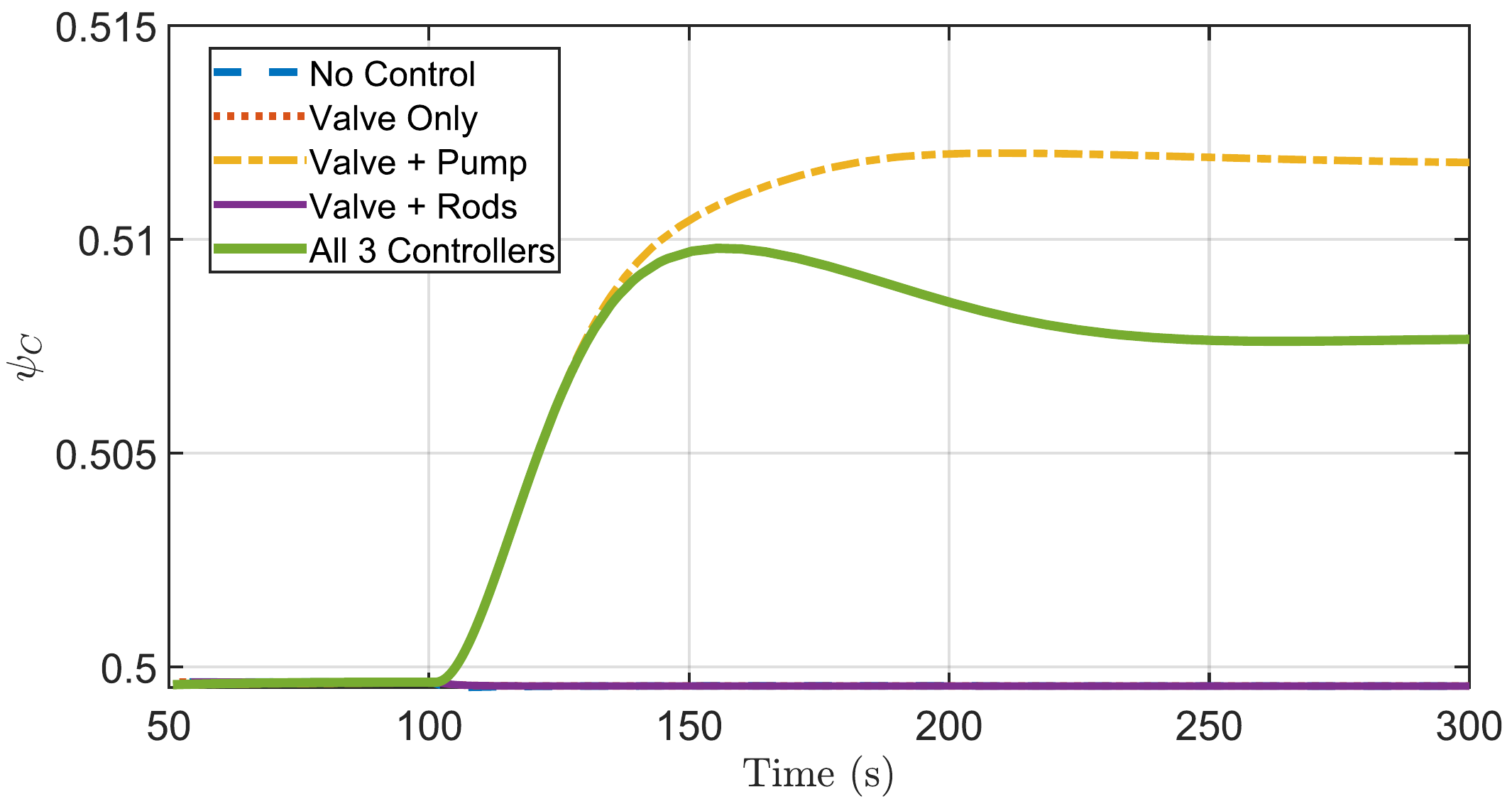}}
  \put(-215,125){\textbf{d}}

  \caption{Steam-generator moving-boundary and cycle-inventory response under different control configurations. The same 5\% step reduction in mechanical-power demand is simulated under Scenarios~1--5. Panels show: \textbf{a,} superheated-region length \(L_3\); \textbf{b,} two-phase-region length \(L_2\); \textbf{c,} subcooled-region length \(L_1\), with \(L_1+L_2+L_3=L_T\); and \textbf{d,} condenser liquid volume fraction \(\psi_C\). Shifts in \(L_1\), \(L_2\), and \(L_3\) show how pressure regulation and feedwater--steam flow mismatch redistribute the secondary-side inventory during the transient.}
  \label{fig:scenario_set2}
\end{figure}

Figures~\ref{fig:scenario_set1} and~\ref{fig:scenario_set2} summarize the main transient responses across the five scenarios. Figure~\ref{fig:scenario_set1} shows the actuator, power, pressure, flow, reactivity, and primary-temperature responses, while Fig.~\ref{fig:scenario_set2} shows the corresponding SG moving-boundary and condenser-inventory behavior.

\subsection{Dynamic response under decentralized control configurations}\label{Sec:scenario-results}

\subsubsection{Natural thermodynamic response}\label{Sec:scenario1}

Scenario~1 provides an open-loop baseline for the coupled plant dynamics. The valve effective area is manually reduced to approximately \(0.021~\mathrm{m^2}\) at \(t=100~\mathrm{s}\) (Fig.~\ref{fig:scenario_set1}a), while the feedwater pump remains in feed-forward mode and the control rods remain fixed. The turbine mechanical power initially drops to approximately \(44~\mathrm{MW}\), but then passively recovers toward the new demand of \(47.5~\mathrm{MW}\) as the SG pressure rises (Fig.~\ref{fig:scenario_set1}b).

This passive recovery is driven by secondary-side pressure accumulation. Because reactor thermal power decreases only slightly, settling near \(152.5~\mathrm{MW}\) (Fig.~\ref{fig:scenario_set1}c), excess energy remains in the SG and raises the secondary pressure from approximately \(3.45~\mathrm{MPa}\) to \(3.70~\mathrm{MPa}\) (Fig.~\ref{fig:scenario_set1}e). The higher upstream pressure increases the driving force across the fixed valve restriction, allowing the steam flow to recover to approximately \(64.2~\mathrm{kg\,s^{-1}}\) (Fig.~\ref{fig:scenario_set1}f). The elevated secondary pressure also increases the saturation temperature, reduces the primary-to-secondary temperature difference, and causes the average primary coolant temperature to settle near \(285.2^{\circ}\mathrm{C}\) (Fig.~\ref{fig:scenario_set1}h). However, the SG moving-boundary and condenser-inventory deviations remain small because the feedwater flow tracks the steam outflow and no sustained inlet--outlet mass-flow mismatch is introduced (Fig.~\ref{fig:scenario_set2}). Thus, the main limitation of Scenario~1 is not SG inventory redistribution, but the passive drift of SG pressure and primary coolant temperature.

\subsubsection{Valve-only power control}\label{Sec:scenario2}

Scenario~2 activates the steam-throttle-valve controller while the feedwater pump remains in feed-forward mode and the control rods remain fixed. The valve controller maintains the turbine mechanical power near \(47.5~\mathrm{MW}\) (Fig.~\ref{fig:scenario_set1}b), eliminating the large transient power sag observed in Scenario~1. However, this improvement in mechanical-power tracking does not restore the SG pressure or the primary thermal state.

Because reactor thermal power decreases only through passive reactivity feedback, the secondary side continues to receive more thermal energy than is removed by the reduced turbine load. The SG pressure again rises to approximately \(3.70~\mathrm{MPa}\) (Fig.~\ref{fig:scenario_set1}e), and the valve constricts to approximately \(0.021~\mathrm{m^2}\) (Fig.~\ref{fig:scenario_set1}a) to maintain the target mechanical power under the elevated upstream pressure. The resulting steam flow remains near \(64.2~\mathrm{kg\,s^{-1}}\) (Fig.~\ref{fig:scenario_set1}f). As in Scenario~1, the SG moving-boundary and condenser-inventory deviations remain comparatively small because the feedwater pump continues to track the steam outflow (Fig.~\ref{fig:scenario_set2}). Therefore, valve-only control is effective for mechanical-power tracking, but it does not recover the nominal thermodynamic state of the SG or primary loop.

\subsubsection{Valve and feedwater-pump control}\label{Sec:scenario3}

Scenario~3 adds SG pressure feedback through the feedwater pump while the control rods remain fixed. The feedwater controller counteracts the initial pressure rise by temporarily reducing the commanded feedwater flow below the steam outflow, with a minimum of approximately \(56.5~\mathrm{kg\,s^{-1}}\) (Fig.~\ref{fig:scenario_set1}g). This mass-flow deficit restores the SG pressure to its nominal value of approximately \(3.45~\mathrm{MPa}\) (Fig.~\ref{fig:scenario_set1}e).

Restoring the secondary pressure changes the valve and flow requirements. With the SG pressure no longer drifting to the high-pressure equilibrium of Scenarios~1 and~2, the valve settles at a wider restriction area of approximately \(0.026~\mathrm{m^2}\) (Fig.~\ref{fig:scenario_set1}a), and the steam flow decreases to approximately \(61.8~\mathrm{kg\,s^{-1}}\) (Fig.~\ref{fig:scenario_set1}f). This indicates reduced secondary-side throttling for the same mechanical-power demand.

However, pressure regulation without active rod control introduces a thermal and moving-boundary trade-off. Because reactor power is adjusted only through passive reactivity feedback, the reduced feedwater flow increases the heat absorbed per unit mass of secondary fluid. As a result, the average primary coolant temperature reaches the highest peak among the five scenarios, approximately \(285.4^{\circ}\mathrm{C}\) (Fig.~\ref{fig:scenario_set1}h). The sharper secondary-side enthalpy rise also reduces the subcooled length \(L_1\) to approximately \(2.6~\mathrm{m}\) (Fig.~\ref{fig:scenario_set2}c), expands the superheated region (Fig.~\ref{fig:scenario_set2}a), and increases the condenser liquid fraction to approximately \(0.512\) (Fig.~\ref{fig:scenario_set2}d). Although the disturbance remains within acceptable operating margins, Scenario~3 shows that pressure control alone can increase primary-temperature excursion and reduce subcooled-region margin when the reactor power is adjusted only passively.

\subsubsection{Valve and control-rod control}\label{Sec:scenario4}

Scenario~4 activates the valve and control-rod controllers while the feedwater pump remains in feed-forward mode. The control rods insert negative reactivity after the load reduction (Fig.~\ref{fig:scenario_set1}d), causing reactor thermal power to drop rapidly to approximately \(149.6~\mathrm{MW}\) before recovering toward a new steady state (Fig.~\ref{fig:scenario_set1}c). This active reactor response limits the primary-temperature excursion compared with Scenario~3, with the average primary coolant temperature peaking near \(284.5^{\circ}\mathrm{C}\) and then returning toward its nominal value (Fig.~\ref{fig:scenario_set1}h).

The active reduction in reactor power also reduces the secondary-pressure rise, but it does not eliminate it. Because the feedwater pump does not use pressure feedback, it cannot correct the pressure deviation caused by changes in secondary-side density and specific volume. The SG pressure therefore settles near \(3.61~\mathrm{MPa}\) (Fig.~\ref{fig:scenario_set1}e), lower than in Scenarios~1 and~2 but still above nominal. The valve consequently remains constricted near \(0.021~\mathrm{m^2}\) (Fig.~\ref{fig:scenario_set1}a), and the SG moving-boundary and condenser-inventory responses remain close to the pressure-unregulated cases (Fig.~\ref{fig:scenario_set2}). Scenario~4 therefore shows that rod control can recover the primary thermal state, but it cannot by itself restore SG pressure or avoid the elevated-pressure throttling behavior associated with feed-forward-only operation.

\subsubsection{Integrated valve--pump--rod control}\label{Sec:scenario5}

Scenario~5 activates all three feedback loops: the valve regulates turbine mechanical power, the feedwater pump regulates SG pressure, and the control rods regulate average primary coolant temperature. The valve controller maintains \(P_{\mathrm{mech}}\) near \(47.5~\mathrm{MW}\) (Fig.~\ref{fig:scenario_set1}b), while the feedwater controller transiently reduces the feedwater flow to counteract the initial pressure rise (Fig.~\ref{fig:scenario_set1}g). As in Scenario~3, this feedwater-flow reduction restores the SG pressure to approximately \(3.45~\mathrm{MPa}\) (Fig.~\ref{fig:scenario_set1}e). At the same time, the control rods reduce reactor thermal power and return the average primary coolant temperature close to its nominal initial value (Fig.~\ref{fig:scenario_set1}c,h).

The coordinated response combines the advantages of Scenarios~3 and~4 while reducing their individual drawbacks. Pressure regulation allows the valve to settle at a wider restriction area of approximately \(0.026~\mathrm{m^2}\) (Fig.~\ref{fig:scenario_set1}a), while the required steam flow decreases to approximately \(62.5~\mathrm{kg\,s^{-1}}\) (Fig.~\ref{fig:scenario_set1}f), compared with approximately \(64.2~\mathrm{kg\,s^{-1}}\) in the valve-only case. This indicates less restrictive valve operation and a lower steam-flow requirement for the same mechanical-power demand.

The moving-boundary response shows that some boundary motion is an inherent consequence of pressure regulation. The temporary mismatch between feedwater inflow and steam outflow redistributes inventory within the SG and condenser, causing the subcooled region to contract to approximately \(2.78~\mathrm{m}\) and the superheated region to expand to approximately \(3.9~\mathrm{m}\) (Fig.~\ref{fig:scenario_set2}a,c). Compared with Scenario~3, however, active rod control prevents the deeper reduction in \(L_1\), and the condenser liquid fraction settles below the value observed in the valve--pump case (Fig.~\ref{fig:scenario_set2}d). Overall, Scenario~5 is the only configuration that simultaneously tracks mechanical-power demand, restores SG pressure, limits primary-temperature drift, and maintains acceptable SG phase-boundary margins.

\subsection{Comparison with a linear transfer-function steam-cycle model}\label{Sec:linear-comparison}

The previous subsection shows that pressure-flow coupling and coordinated control strongly affect the transient response of the physics-based SMR--Rankine model. To quantify the effect of the secondary-cycle representation itself, the proposed physics-based framework was compared with a simplified linear transfer-function baseline. The same equation-based SMR and SG model and the same integrated valve--pump--rod controller were used in both cases. The only difference is the representation of the secondary Rankine cycle. The parameters of the reduced baseline were selected so that both models reproduce the same rated operating point before the load-following disturbance.

\begin{figure}[thbp!]
  \centering
  \includegraphics[width=\textwidth]{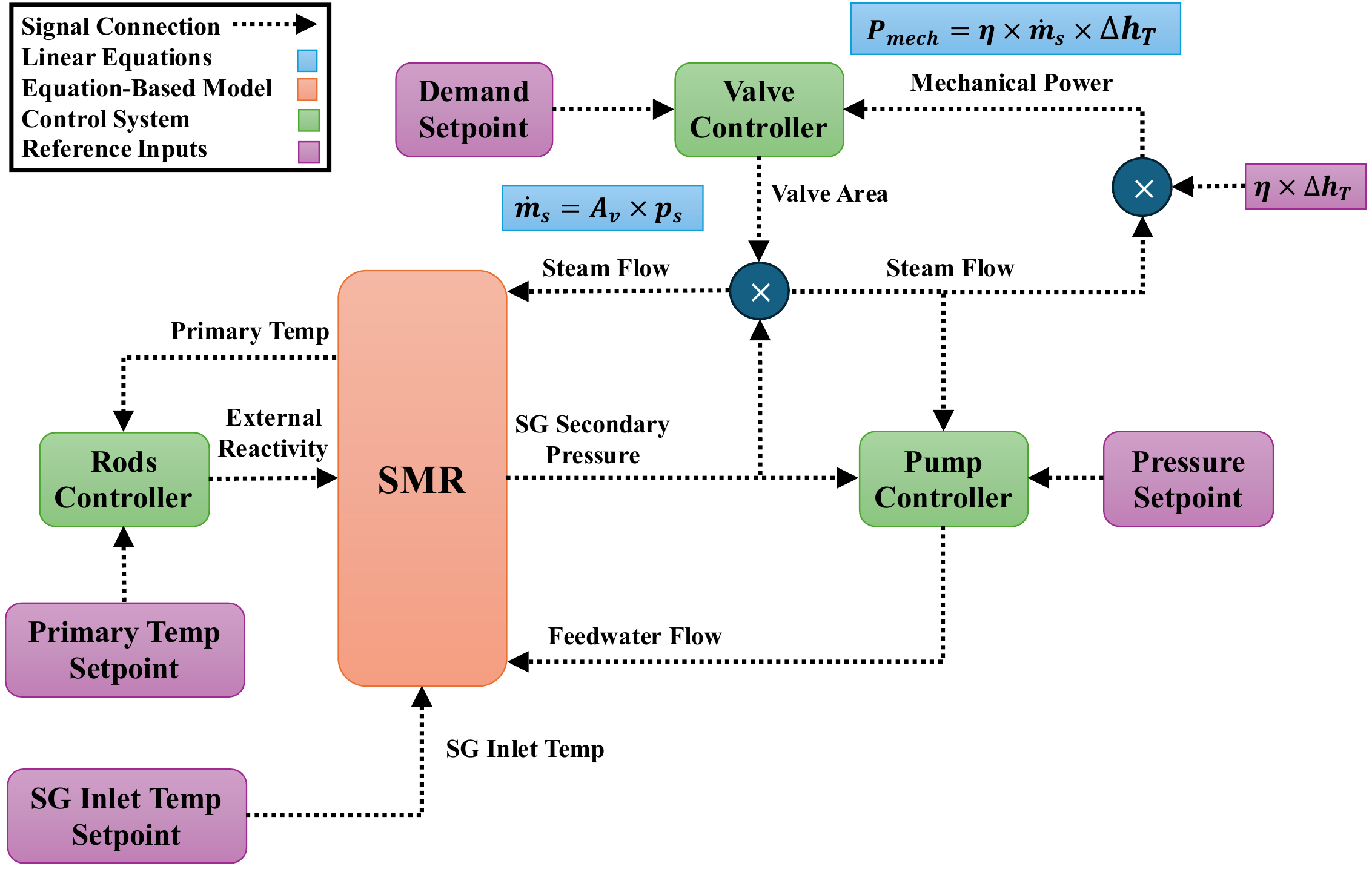}
  \caption{Linear transfer-function baseline architecture. The equation-based SMR and SG model and the decentralized control structure are retained, but the Rankine-cycle balance of plant is reduced to causal signal-flow blocks. Steam flow is prescribed from simplified valve and pressure relations, and turbine mechanical power is computed from steam flow using a fixed effective specific work. As a result, downstream two-phase coupling, condenser inventory, and state-dependent turbine expansion are not resolved explicitly.}
  \label{fig:Block_diagram_Linear_model}
\end{figure}

In the linear baseline, the Rankine-cycle balance of plant is represented as a causal signal-flow network. The steam flow is prescribed from simplified algebraic or transfer-function relations, and turbine mechanical power is computed from steam flow using a fixed effective specific work. This structure is representative of control-oriented models in which the secondary-cycle dynamics are compressed into gains, time constants, and actuator limits \cite{vajpayee2020dynamic}. Figure~\ref{fig:Block_diagram_Linear_model} summarizes the baseline architecture. Unlike the physics-based model, the linear baseline does not include acausal two-phase fluid ports, dynamic condenser inventory, two-phase pressure-dependent valve flow, or load-dependent turbine enthalpy drop.

For clarity, the reduced valve and turbine relations used in the linear baseline can be represented as
\begin{equation}
\dot{m}_s(t)
=
K_{V,\mathrm{lin}} A_V^R(t)p_s(t),
\label{eq:linear_valve}
\end{equation}
and
\begin{equation}
P_{\mathrm{mech}}(t)
=
\eta_{\mathrm{mech}}\dot{m}_s(t)\Delta h_0,
\label{eq:linear_turbine}
\end{equation}
where \(K_{V,\mathrm{lin}}\) is an effective linear valve-flow coefficient, \(A_V^R(t)\) is the effective valve restriction area, \(p_s(t)\) is the SG pressure, and \(\Delta h_0\) is a constant nominal turbine specific enthalpy drop. The coefficient \(K_{V,\mathrm{lin}}\) is used only in the reduced baseline model and should not be confused with the pressure-dependent resistance factor \(K_V(\Delta p_V)\) used in the physics-based valve model. These simplified relations contrast with the physics-based valve and turbine models, in which steam flow depends on the instantaneous pressure drop and fluid state, and the turbine enthalpy drop is recalculated from the current inlet and outlet thermodynamic states.

\subsubsection{Comparative transient response under integrated control}\label{Sec:linear-transient}

\begin{figure}[thbp!]
  \centering
  
  \subfloat{\includegraphics[width=0.48\textwidth]{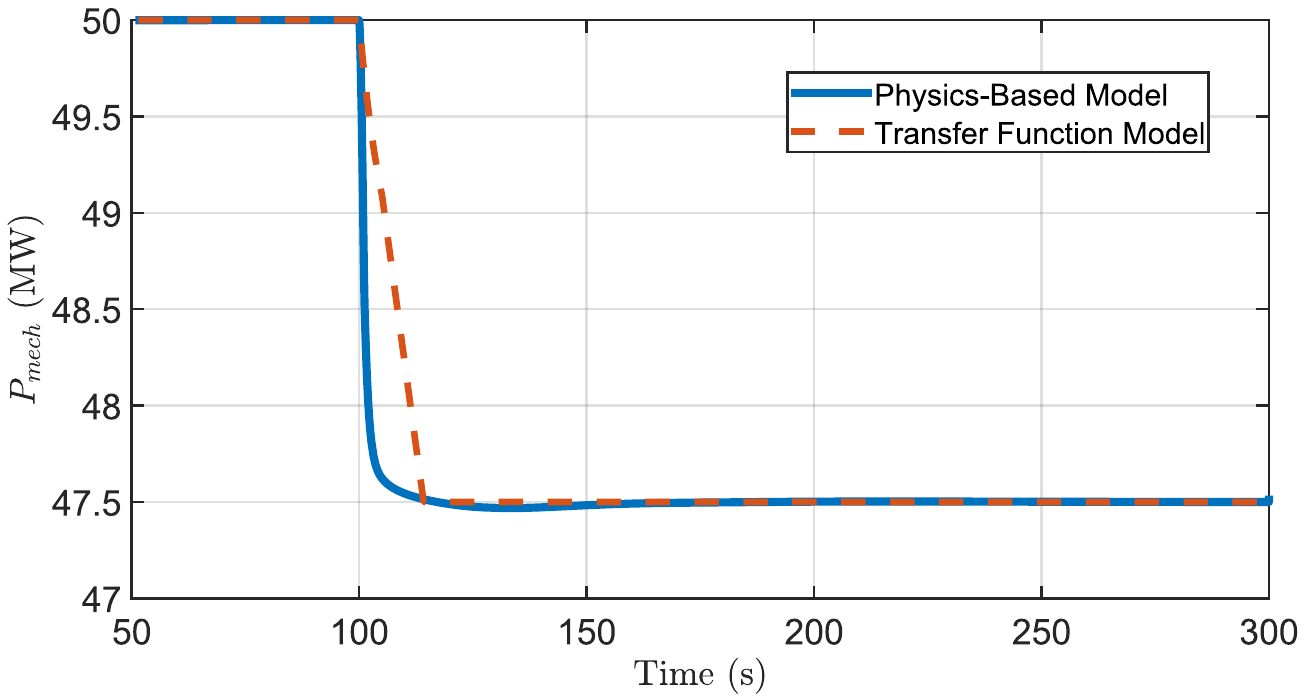}}
  \put(-215,125){\textbf{a}}
  \hspace{1em}
  \subfloat{\includegraphics[width=0.48\textwidth]{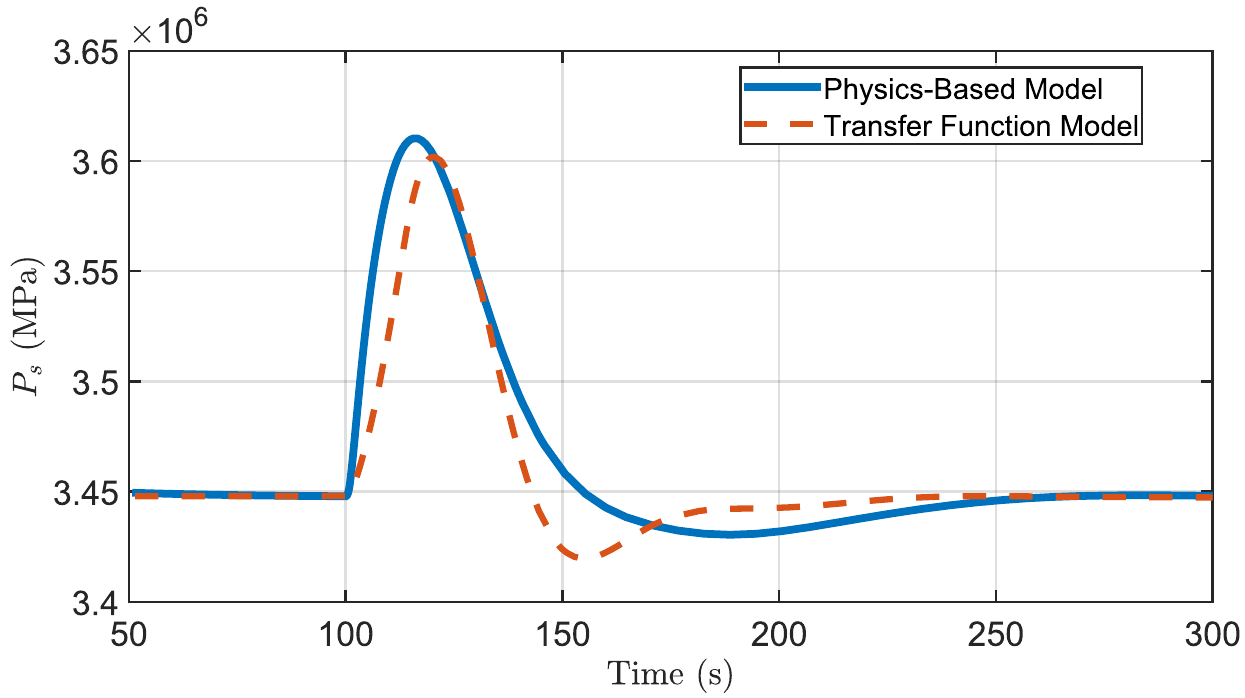}}
  \put(-215,125){\textbf{b}}
  
  \subfloat{\includegraphics[width=0.48\textwidth]{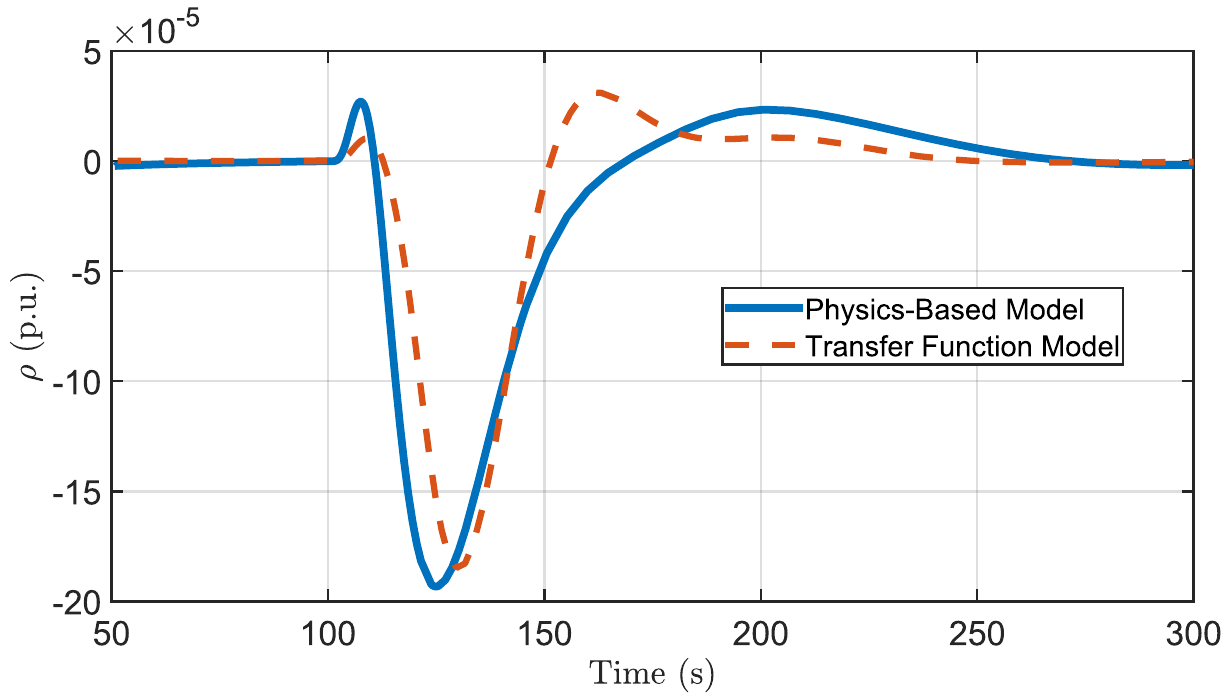}}
  \put(-215,125){\textbf{c}}
  \hspace{1em}
  \subfloat{\includegraphics[width=0.48\textwidth]{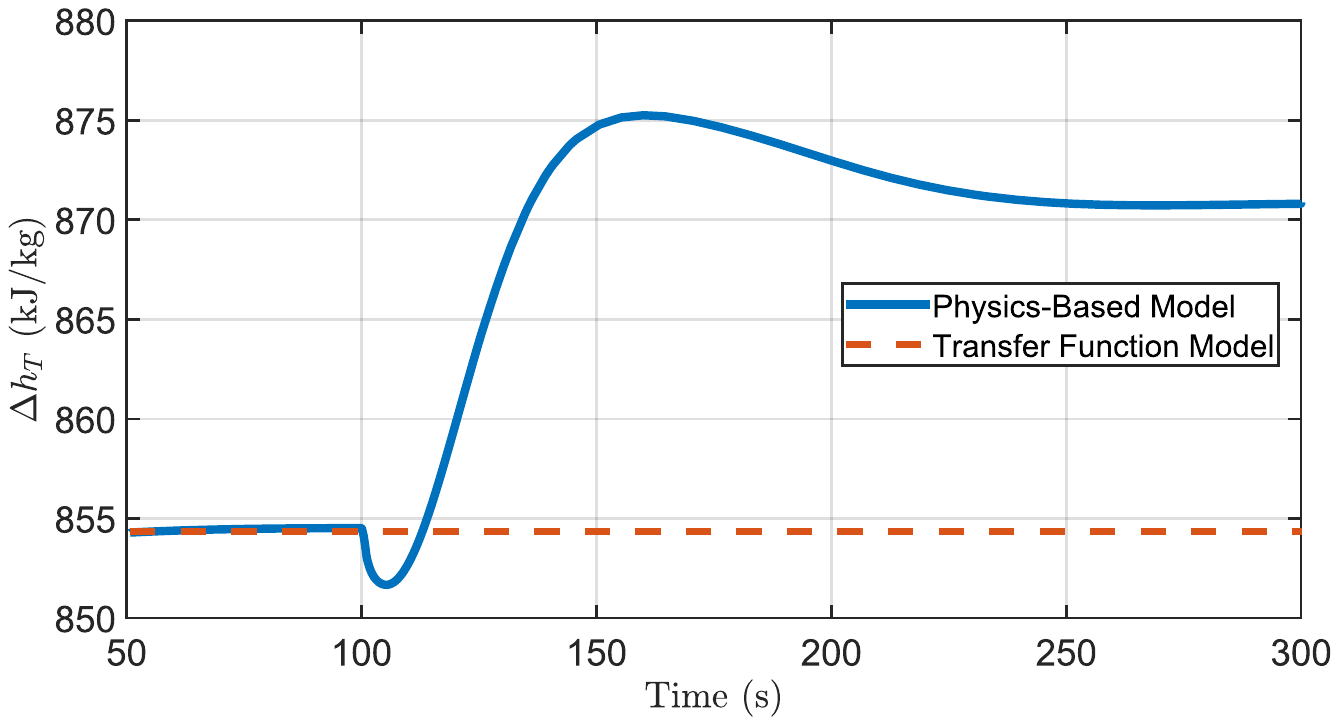}}
  \put(-215,125){\textbf{d}}
  
  \subfloat{\includegraphics[width=0.48\textwidth]{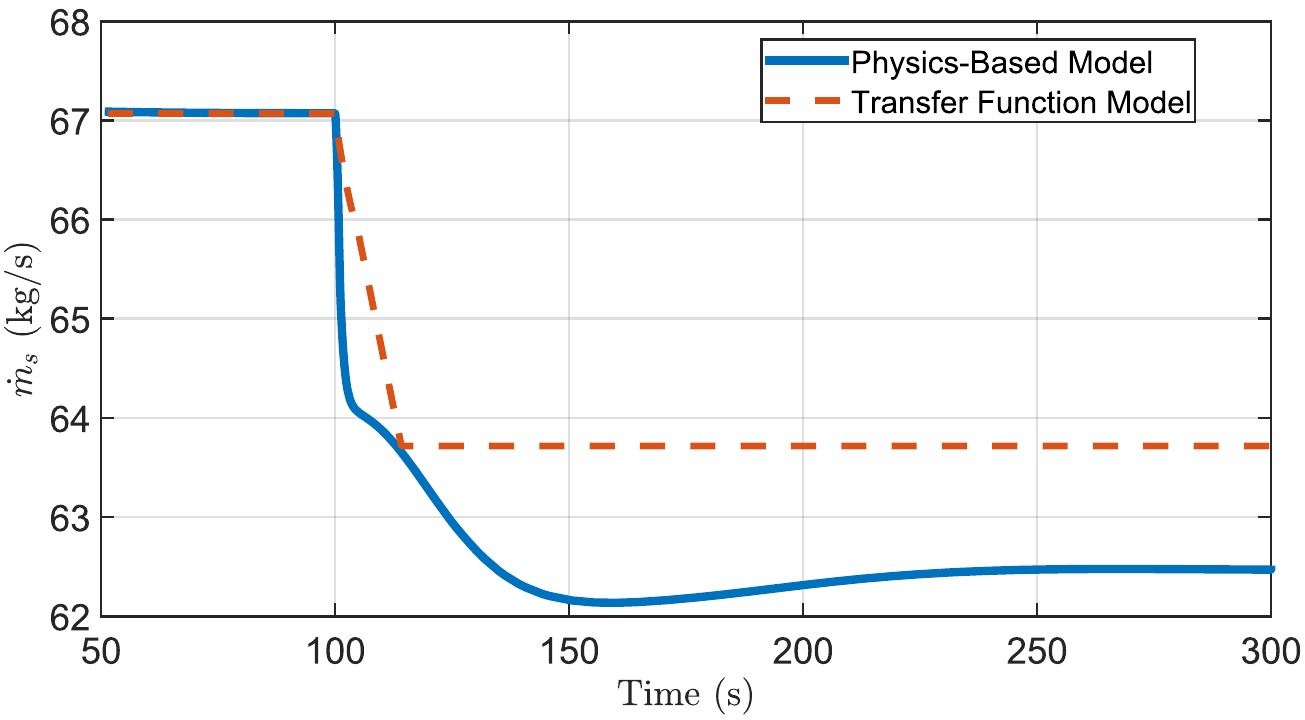}}
  \put(-215,125){\textbf{e}}
  \hspace{1em}
  \subfloat{\includegraphics[width=0.48\textwidth]{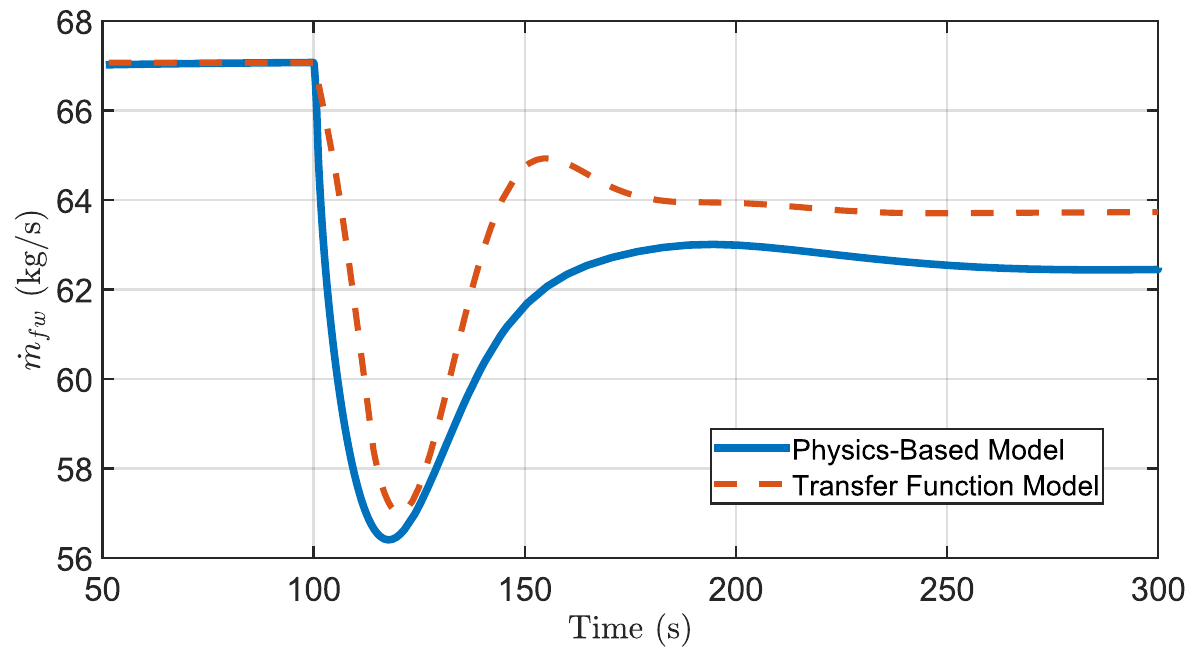}}
  \put(-215,125){\textbf{f}}
  
  \subfloat{\includegraphics[width=0.48\textwidth]{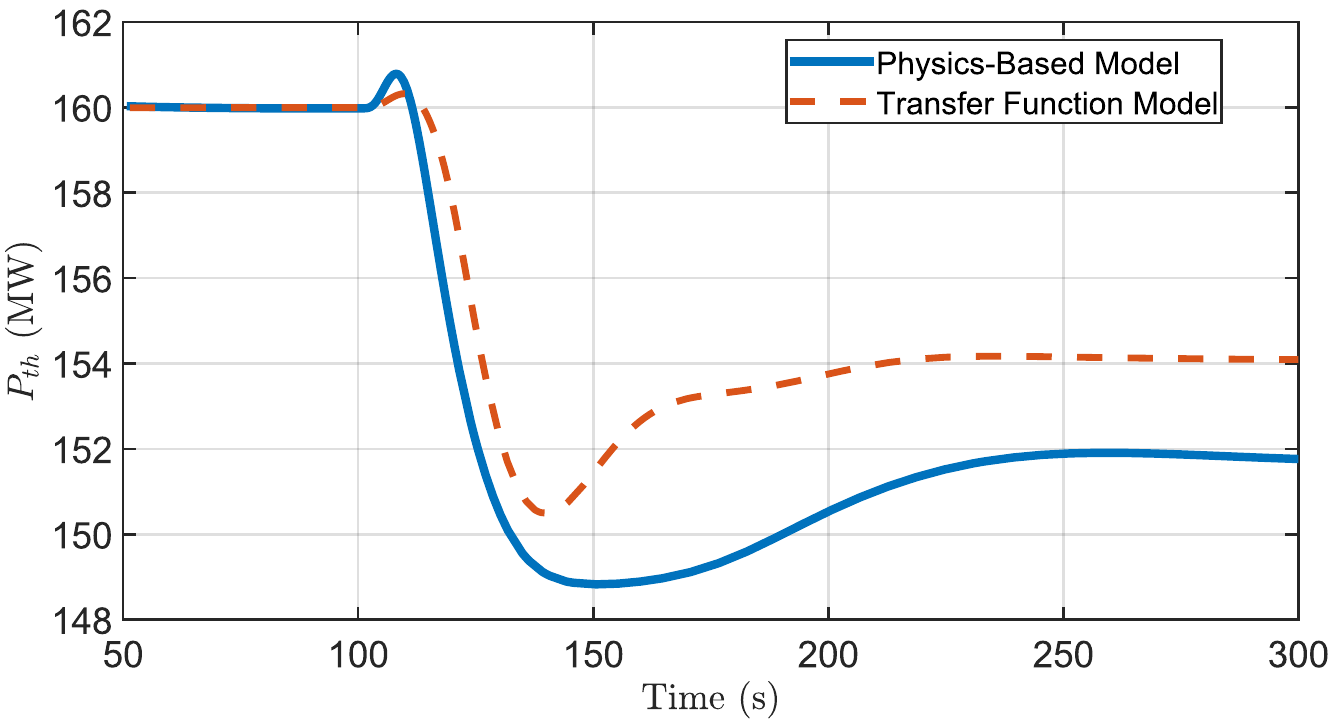}}
  \put(-215,130){\textbf{g}}
  \hspace{1em}
  \subfloat{\includegraphics[width=0.48\textwidth]{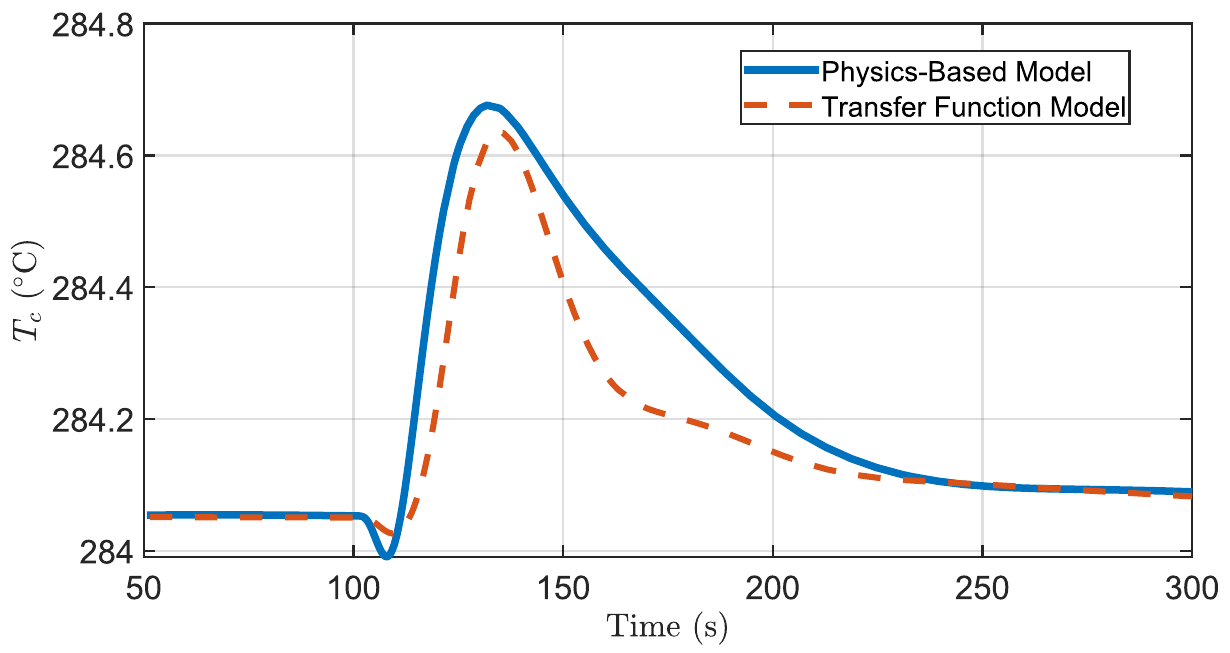}}
  \put(-215,125){\textbf{h}}
  
  \caption{Comparison of the physics-based and linear transfer-function models during near-nominal load following. Both models are subjected to the same 5\% step reduction in mechanical-power demand, from \(50~\mathrm{MW}\) to \(47.5~\mathrm{MW}\), at \(t=100~\mathrm{s}\), using the same SMR formulation and integrated control structure. Panels show: \textbf{a,} turbine mechanical power \(P_{\mathrm{mech}}\); \textbf{b,} SG secondary pressure \(p_s\); \textbf{c,} core reactivity \(\rho_{\mathrm{react}}\); \textbf{d,} specific enthalpy drop across the turbine \(\Delta h_T\); \textbf{e,} steam mass flow rate \(\dot{m}_s\); \textbf{f,} feedwater mass flow rate \(\dot{m}_{\mathrm{fw}}\); \textbf{g,} reactor thermal power \(P_{\mathrm{th}}\); and \textbf{h,} primary coolant average temperature \(\bar{T}_c\).}
  \label{fig:compare_set1}
\end{figure}

Both models were subjected to the same 5\% step reduction in mechanical-power demand under the integrated valve--pump--rod control configuration. Figure~\ref{fig:compare_set1} compares the main plant responses. Both models track the new mechanical-power demand, confirming that the control objective is satisfied in both representations (Fig.~\ref{fig:compare_set1}a). Both models also predict an initial SG pressure rise after the valve begins to close (Fig.~\ref{fig:compare_set1}b). However, the pressure recovery differs: the linear model exhibits a sharper undershoot below the nominal pressure, whereas the physics-based model shows a more damped recovery. This difference arises because the physics-based model resolves pressure-flow coupling through the valve, turbine, condenser, and feedwater network rather than imposing the response through fitted lags.

The reactivity and primary-temperature responses also differ after the initial transient (Fig.~\ref{fig:compare_set1}c,h). Although both models insert negative reactivity to reduce reactor thermal power after the load reduction, the physics-based model produces a broader recovery because the secondary-side pressure--flow response modifies the heat-removal dynamics imposed on the primary loop.

\subsubsection{Thermodynamic-gap effects}\label{Sec:thermodynamic-gap}

The most important difference between the two models appears in the secondary-cycle thermodynamic variables. In the linear baseline, the turbine specific enthalpy drop remains fixed at the nominal value by construction. In the physics-based model, \(\Delta h_T\) is recalculated from the instantaneous turbine inlet and outlet states and therefore changes with pressure ratio and load condition. As shown in Fig.~\ref{fig:compare_set1}d, the physics-based model predicts an increase of approximately 1.8\% in the turbine specific enthalpy drop as the system settles to the new part-load equilibrium.

This load-dependent turbine specific work changes the mass and energy balance of the entire plant. Because the physics-based turbine extracts more mechanical work per unit mass of steam at the new operating point, less steam is required to produce the same \(47.5~\mathrm{MW}\) output. The physics-based model settles at a lower steady-state steam flow, approximately \(62.5~\mathrm{kg\,s^{-1}}\), than the linear baseline (Fig.~\ref{fig:compare_set1}e). The feedwater flow follows the same trend because the pump controller restores SG inventory balance at the new operating point (Fig.~\ref{fig:compare_set1}f).

The lower secondary-side mass flow also reduces the thermal power required from the reactor. As shown in Fig.~\ref{fig:compare_set1}g, the physics-based model settles near \(152~\mathrm{MW}\), whereas the linear model requires more than \(154~\mathrm{MW}\) to produce the same mechanical output. This difference quantifies the thermodynamic gap introduced by simplified transfer-function representations: by assuming a constant turbine enthalpy drop and suppressing pressure-flow coupling, the linear model overpredicts the steam-flow and reactor-power requirements at the part-load condition.

\begin{figure}[tbp!]
  \centering
  
  \subfloat{\includegraphics[width=0.48\textwidth]{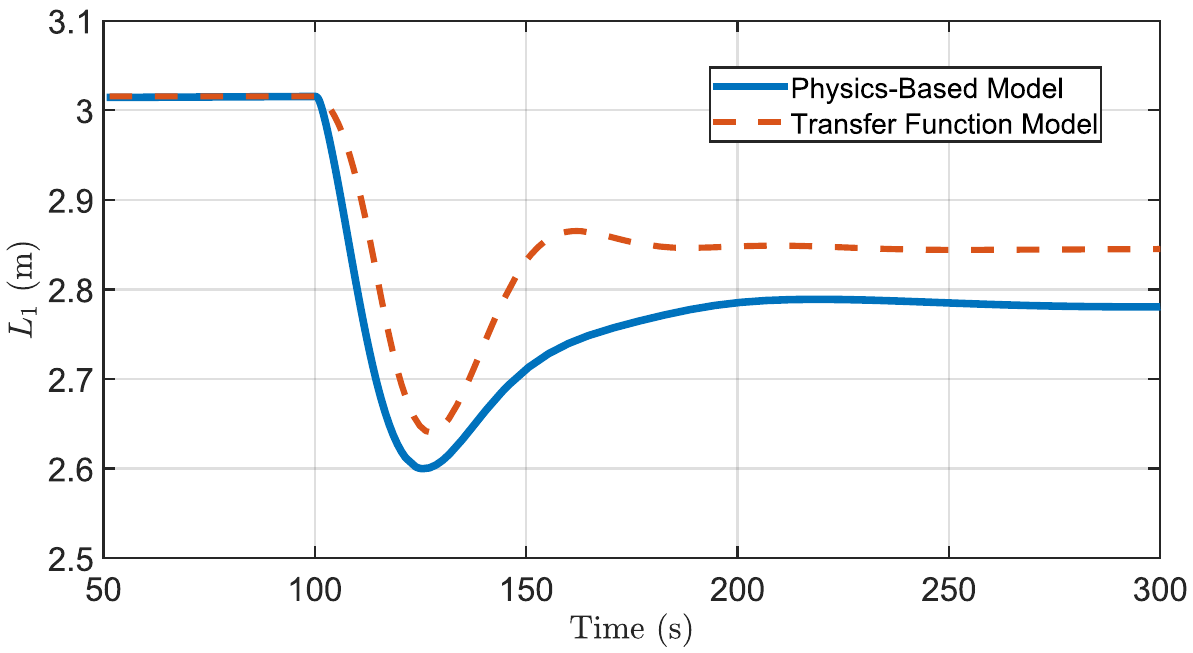}}
  \put(-215,125){\textbf{a}}
  \hspace{1em}
  \subfloat{\includegraphics[width=0.48\textwidth]{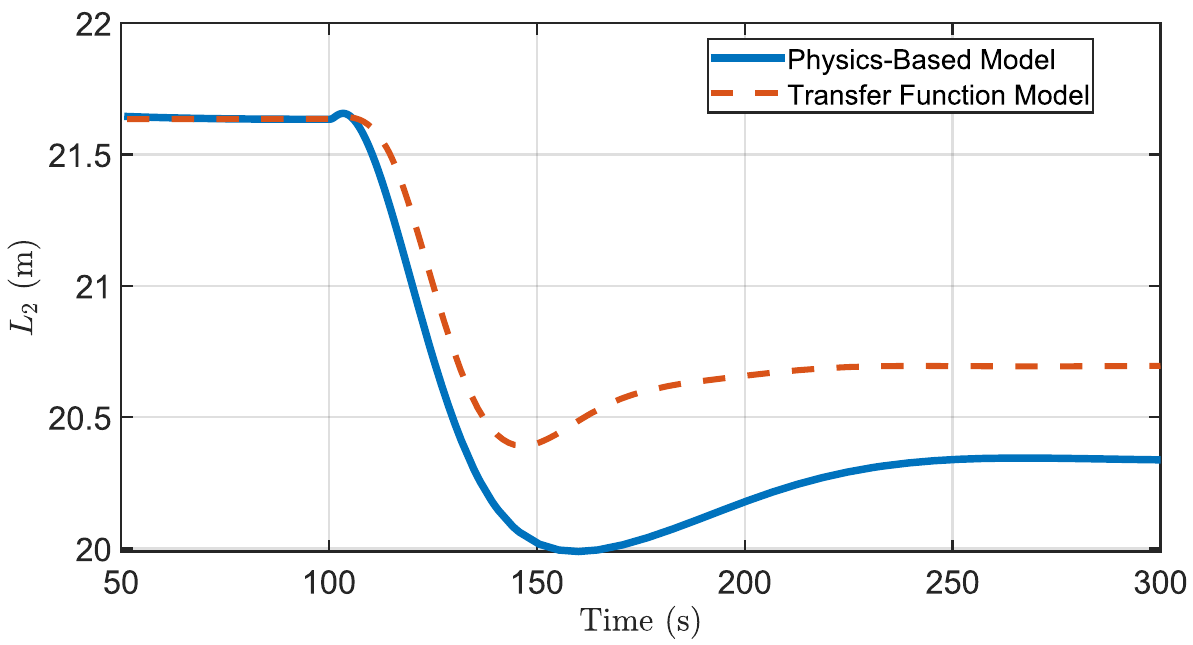}}
  \put(-215,125){\textbf{b}}

  \subfloat{\includegraphics[width=0.48\textwidth]{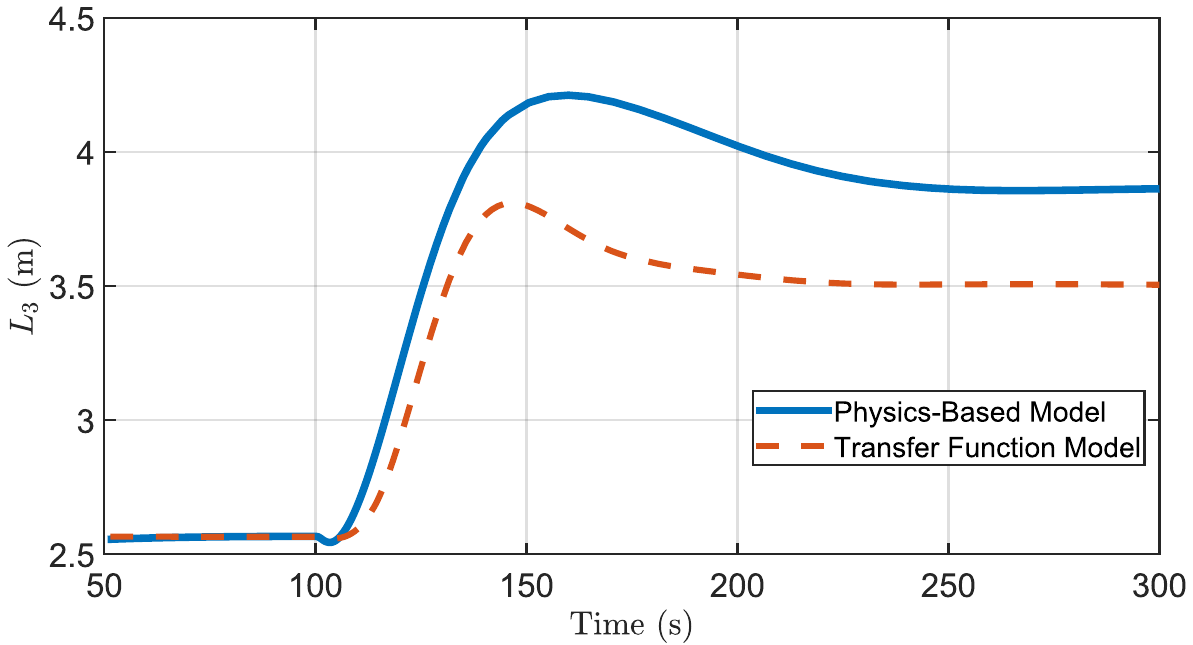}}
  \put(-215,125){\textbf{c}}

  \caption{Thermodynamic-gap effects on SG moving-boundary dynamics. Both the physics-based model and the linear transfer-function baseline are subjected to the same 5\% load reduction using the same SMR formulation and integrated controller. Panels show: \textbf{a,} subcooled-region length \(L_1\); \textbf{b,} two-phase-region length \(L_2\); and \textbf{c,} superheated-region length \(L_3\). Differences arise from the secondary-cycle representation and the resulting mass-flow and enthalpy boundary conditions imposed on the moving-boundary SG.}
  \label{fig:compare_set2}
\end{figure}

The same thermodynamic gap appears in the SG moving-boundary response. Figure~\ref{fig:compare_set2} compares the subcooled, two-phase, and superheated region lengths for the physics-based and linear models. Both models use the same moving-boundary SG equations; therefore, the differences arise from the secondary-side boundary conditions imposed by the steam-cycle representation. Because the physics-based model settles at a lower steam and feedwater flow for the same mechanical output, the secondary fluid absorbs more heat per unit mass. Consequently, the subcooled and two-phase regions contract more than in the linear baseline, while the superheated region expands more significantly. The linear model underestimates these boundary shifts because the constant-specific-work assumption constrains the steam-flow response. These results show that the secondary-cycle representation affects not only turbine and flow predictions but also the internal thermal distribution of the moving-boundary SG. Thus, even when both models achieve nearly identical mechanical-power tracking, they predict different internal thermodynamic states, demonstrating that matching the external power response alone is insufficient for assessing SMR load-following behavior and SG operating margins.

\section{Conclusion}\label{Sec:conclusion}

This study developed a thermodynamically coupled dynamic framework for load-following analysis of an iPWR-based SMR with a moving-boundary helical-coil once-through SG and a physics-based secondary Rankine cycle. The equation-based SMR--SG model was coupled to two-phase Simscape components representing the steam throttle valve, turbine, condenser, and feedwater pump. This structure preserves the SG moving-boundary formulation while allowing secondary pressure, steam flow, condenser inventory, and turbine enthalpy drop to evolve from the coupled thermodynamic state of the plant.

The integrated model reproduced the nominal design-point conditions and was used to evaluate a 5\% step reduction in turbine mechanical-power demand under five control configurations. The results show that SMR load-following performance is not determined by turbine-power tracking alone. Valve-only control can regulate mechanical power but leaves SG pressure and primary coolant temperature displaced from their nominal values. Feedwater-pump pressure control restores SG pressure, but without active rod control it increases the primary-temperature excursion and produces larger SG phase-boundary shifts. Rod control limits primary thermal deviation but does not restore the secondary pressure. The integrated valve--pump--rod control configuration provides the most balanced response by tracking mechanical-power demand, restoring SG pressure, limiting primary-temperature drift, and maintaining acceptable SG phase-boundary margins.

Comparison with a simplified linear transfer-function BOP model further demonstrates the importance of the secondary-cycle representation. Although the linear and physics-based models can satisfy similar mechanical-power control objectives, they predict different part-load thermodynamic states. In particular, the physics-based model captures dynamic back-pressure and load-dependent turbine specific enthalpy drop, leading to different predictions of steam-flow requirement, feedwater-flow requirement, reactor thermal power, and SG phase-boundary evolution. These differences quantify the thermodynamic gap introduced when the Rankine cycle is reduced to fixed gains, time constants, or constant-specific-work assumptions.

The present study is limited to a representative near-nominal 5\% load-reduction case and uses a compact reactor and primary-loop model to preserve plant-level computational tractability. Detailed transient validation is also limited by the lack of publicly available dynamic data for the target SMR design. Future work should extend the framework to larger load changes, ramp-following profiles, repeated cycling, condenser-pressure disturbances, and feedwater-temperature perturbations. Additional control studies could also build on the decentralized reactor-following-turbine architecture by explicitly considering SG phase-boundary margins, pressure limits, actuator constraints, and primary thermal excursions. Coupling the proposed SMR--Rankine framework with generator, grid, or hybrid-energy-system models would further enable assessment of SMR flexibility in realistic power-system applications.

Overall, the results show that thermodynamic coupling between the reactor, moving-boundary SG, and secondary Rankine cycle materially affects conclusions about SMR load-following behavior. Physics-based BOP models therefore provide a more physically informative basis for evaluating SMR operational flexibility, control requirements, and operating margins than weakly coupled or purely transfer-function-based representations.

\section*{Declaration of competing interest}
The authors declare that they have no known competing financial interests or personal
relationships that could have appeared to influence the work reported in this paper.

\section*{Acknowledgment}
This work was supported by the U.S. Department of Energy under Award DE-NE0009296.

\section*{Data and code availability}
The source data underlying the figures and tables generated in this study and the code used to produce the results are publicly available at: \href{https://github.com/Ali-Mahboub-Rad/iPWR-SMR-Dynamic-Model}{\textcolor{blue}{https://github.com/Ali-Mahboub-Rad/iPWR-SMR-Dynamic-Model}}. 




\bibliographystyle{elsarticle-num}

\bibliography{references}

\end{document}